\documentclass[preprint,12pt]{elsarticle}
\usepackage{hyperref}
\journal{}

\usepackage{graphicx, color, transparent}
\usepackage{euscript}
\usepackage{amsmath, amsfonts}
\usepackage{breqn}
\usepackage{setspace}
\usepackage{bm}
\usepackage[dvipsnames]{xcolor}
\usepackage{siunitx}
\usepackage{xcolor}
\usepackage{graphicx}
\usepackage{caption}
\usepackage{subcaption}
\usepackage{wasysym}
\usepackage[thinc]{esdiff}

\usepackage[normalem]{ulem}

\newcommand{\rv}[1]{\textcolor{black}{#1}}

\begin{document}

\begin{frontmatter}

\title{Toughening mechanisms and damage propagation in Architected-Interfaces} %\\ Mechanical metamaterial leads to fail-safe interface damage propagation}
%\\Mechanical metamaterials are fracture metamaterials}

\author[1]{Michelle L. S. Hedvard}

\author[2]{Marcelo A. Dias\corref{cor}}
\ead{marcelo.dias@ed.ac.uk}

\author[1]{Michal K. Budzik\corref{cor}}
\ead{mibu@mpe.au.dk}

\cortext[cor]{Corresponding authors}

\address[1]{Department of Mechanical and Production Engineering, Aarhus University, 8000~Aarhus~C, Denmark}
\address[2]{Institute for Infrastructure \& Environment, School of Engineering, The University of Edinburgh, Edinburgh~EH9~3FG Scotland, UK}
%\address[3]{Chair of Lightweight Design, University of Rostock, Germany}
%%%%%%%%%%%%%%%%
\begin{abstract}
%%%%%%%%%%%%%%%%
%We investigate fracture properties of architected interfaces and their ability to ensure structural integrity and stable damage propagation conditions beyond the failure load. The theoretical and numerical frameworks are proposed to evaluate the fracture properties of architected interfaces sandwiched between two (face) materials. The microscopic geometries of these interfaces is chosen as 2D cells---pillar, tetrahedron, and the hexagonal---as well as their 3D relatives---i.e. pillar array, octet truss and Kelvin cell. Our model, both numerical and analytical, shows good mutual accuracy in predicting the pre-failure compliance and failure loads. Novel results are recorded during the damage propagation regime and indicate fulfilment of the so-called fail-safe design. Some of the cell geometries, unfold during fracture increasing the failure load and ensuring stable and controlled damage propagation conditions. 
We investigate fracture \rv{toughness} of architected interfaces and their ability to maintain structural integrity and provide stable damage propagation conditions beyond the failure load. We propose theoretical and numerical frameworks to evaluate the fracture properties of architected interfaces sandwiched between two (face) materials. The microscopic geometries of these interfaces are chosen as 2D cells---pillar, tetrahedron, and hexagon---as well as their 3D counterparts---namely, pillar array, octet truss, and Kelvin cell. Our model, both numerical and analytical, exhibits a high level of accuracy in predicting the compliance before failure and failure loads. Novel results are obtained during the damage propagation regime, indicating fulfilment of the so-called fail-safe design. Some of the cell geometries unfold during fracture, thus increasing the failure load and ensuring stable and controlled damage propagation conditions.
%Depending on the unit cell geometry, the architected interfaces are fracture metamaterials.
\end{abstract}


\end{frontmatter}

%\linenumbers
\newcommand{\mad}[1]{\textcolor{red}{\emph{[MAD: ---#1]}}}
\newcommand{\mibu}[1]{\textcolor{magenta}{\emph{[MIB: ---#1]}}}
\newcommand{\mlsh}[1]{\textcolor{purple}{\emph{[AFA: ---#1]}}}
\definecolor{pk1}{rgb}{.88, 0.36, 0.50}
%------------------
\section{Introduction}
%%%%%%%%%%%%%%%%%%%%%%
Recent advances in manufacturing have enabled fabrication of ever more complex architected materials, which has brought large interest from scientific and industrial communities to advance the future functional materials~\cite{xia2022responsive}. Although architected materials, also \rv{referred to as Mechanical Metamaterials (MMs)~\cite{valdevit2018architected}}, seem promising to innovate mechanical design, such as performance weighted against the effective use of resources~\cite{zheng2014ultralight}, their applications are still limited and focused on compressive and impact loading~\cite{shan2015multistable,chen2020light}. When it comes to damage resistance, one important feature of such materials is a multitude of failure modes depending on the choice of cell geometries and/or direction of applied load. For instance, if the cell symmetry line is aligned with the applied load during compressive loading, cell features localise stresses leading to, for instance, buckling which subsequently create so-called kink bands---such process is often similar to the one known from cellular solids~\cite{mccormack2001failure,rayneau2016recipes,oster2021reentrant}. The failure is energetically expensive but often not critical from the structural integrity perspective~\cite{injeti2019metamaterials}. However, once failure occurs under the tensile loading, symmetry considerations are a key concept: a lattice structure with its local (\emph{i.e.} cell) symmetry plane aligned with the globally applied loading conditions lacks capability of redistributing the stresses and the instantaneous excess of stored energy leads to spontaneous failure. Such failure under tensile loading have been associated with resistance against fracture~\cite{shaikeea2022toughness} in the hopes that fracture mechanics can enable better future designs. A number of studies focused on defining effective stress intensity factors and successfully deduce effective fracture properties of MMs~\cite{quintana2009fracture,shaikeea2022toughness,omidi2023fracture}. 
One of the potential applications of MMs is adhesive bonding~\cite{athanasiadis2021confined}, with the notion that for wide use of MMs, they will need to be integrated into existing and future designs. In adhesive bonding, or integrated scenario, MMs will be confined between two layers. We shall refer to such use of MMs as Architected-Interfaces (ArchIs).

The verge of adhesive bonding is associated with aerospace engineering and up until this point in the history of adhesive bonding, the majority of aerospace applications adhesive layers are kept thin, ca. $100-300\,{\mu}$m. However, the expansion of bonding to wind turbine blades, maritime structures or civil engineering requires substantially different look at adhesive layer when compared against those adopted in aerospace joints. Most of the time, in the aforementioned applications, joints are designed to bond dissimilar materials and manufacturing tolerances as well as on-site conditions are usually far from controllable laboratory conditions characteristic for aerospace. Moreover, physical, chemical, and mechanical requirements for bonding materials are far different from that in aerospace. Here, the preference seems to be given to a more flexible but tougher materials. It is noteworthy that the micrometer bondline thicknesses are never used, instead they vary from millimeters to centimeters. For instance, $20-30$ mm thick bondlines are used for joining wind-turbine blades~\cite{Review_of_fatigue_of_bulk_structural_adhesives,Analysis_and_evaluation_of_bondline_thickness_effects}.
Many challenges present themselves in the context of bonding of large engineering structures. With added flexibility and toughness, the volume of adhesive used is substantially increased, while adhesives are usually not regarded as environmentally friendly---although still new, green and recyclable compositions are being developed. Therefore, it seems reasonable to claim, that what can be a solution is `high volume'-to-`low mass' bondline layers. This is indeed the very core of ArchIs working principle~\cite{athanasiadis2021confined}.

The mechanical properties of the bondlines are also affected by the thickness. If the adhesive material is elastic-brittle, corner singularities will lead to interface crack onset and premature failure. Therefore, ductile materials are preferred for thick bondlines with advantage of higher fracture toughness and increased damping capabilities \cite{The_effect_of_bond-line_thickness_on_fatigue_crack_growth_rate,The_Effect_of_Adhesive_Thickness_on_the_Mechanical_Behavior}. Such capabilities are within the reach of ArchIs, where properties can be adjusted through geometrical design. However, the full understanding of ArchIs confined between the joined materials, affecting, distribution of cells as well as global and local symmetries under loading is currently lacking.
In addition, as adhesive joints are multi-scale structures, ArchIs will add at least two further scales---one of a cell size and the other from size of the most fundamental structural building blocks, i.e. trusses, plates or shells. In that respect, modelling and evaluation of metamaterial-inspired bondlines require development. In recent work~\cite{athanasiadis2021confined}, a theoretical analysis to compute failure loads for simpler 2D lattice systems mimicking bondlines was proposed lead to establishing relations between the geometry and the load at damage initiation. In another development~\cite{shaikeea2022toughness}, the damage onset was investigated experimentally through tomography technique. 
%However, no data is available from the damage propagation stage.
\rv{Some studies, such as those exploring numerical methods for toughening in 2D lattice materials and investigating crack growth in elastoplastic lattices for enhanced toughness through multi-phase reinforcement strategies~\cite{hsieh2020versatile,tankasala2020crack}, have focused on revealing the architecture and material impact on fracture behaviour. However, still limited research has been conducted on the damage propagation stage in architected materials, especially for those of confined nature.}
Therefore, the focus in this work is two fold. Primarily we wish to expand metamaterial-inspired bondlines concept by introducing a numerical framework for studying these structures in 3-dimensions, with focus on post failure-onset behaviour. The second focus is on witnessed fracture toughness increase during the crack growth that can be harnessed as safe-fail design concepts. We shall focus on the the Double Cantilever Beam (DCB) setup for the numerical experiments, where ArchIs are used as beam-joining interfaces. 

%%%%%%%%%%%%%%%%%%%%%%%%%%%%%%%%%%%
\section{Theoretical analysis of architected interfaces}
%%%%%%%%%%%%%%%%%%%%

\begin{figure}[!h]
    \centering
    \includegraphics[height=5.5cm]{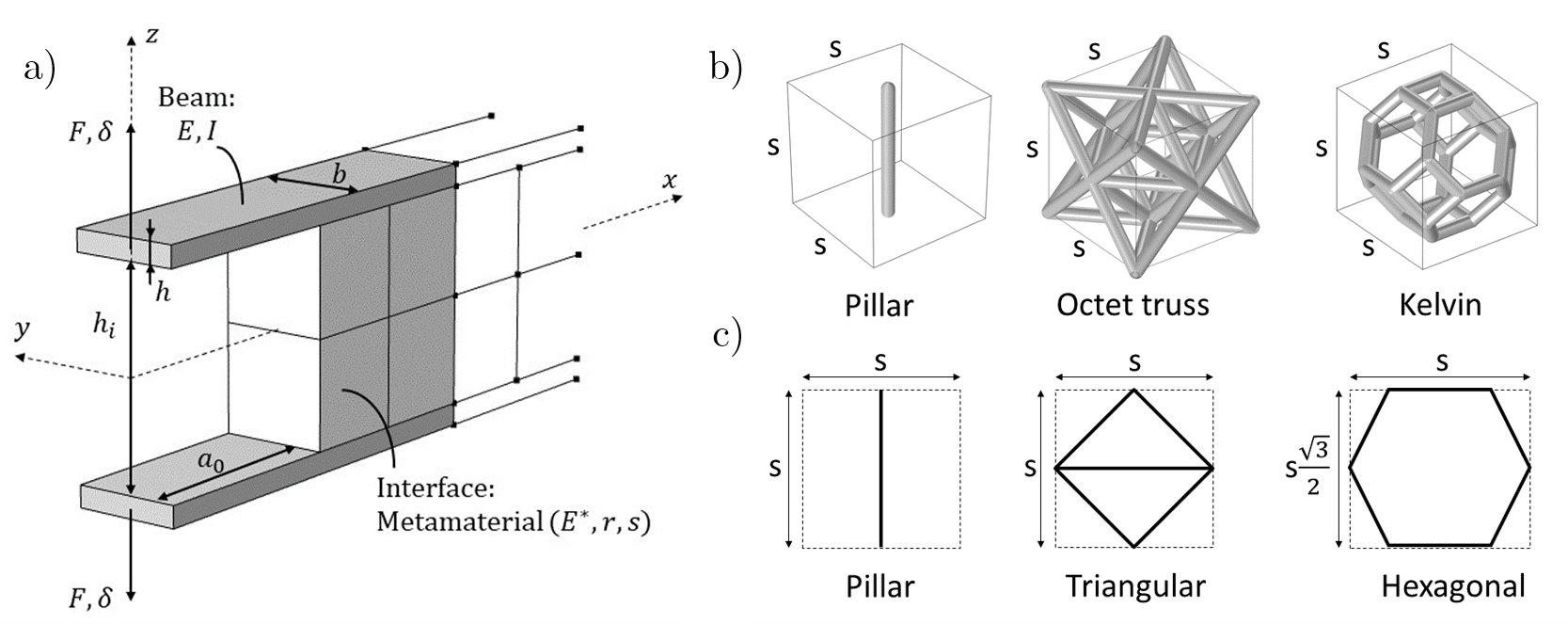}
    \caption{(a) Schematic representation of the DCB setup. (b) 3D cells, the dimensions of the cells are $s\times s\times s$. (c) 2D cells, the dimensions are given in the figure.}
    \label{fig:Scope}
\end{figure}

We study damage propagation in ArchIs in 2-dimensions (2D) and 3-dimensions (3D). 2D interfaces are constructed from pillar, triangular, and hexagonal cells, whereas in 3D, pillar array, octet truss, and Kelvin cells are used---these geometries are shown in Fig.~\ref{fig:Scope}.
%%%%%%%%%%%%%%%%%%%%
%%%%%%%%%%%%%%%%%%%%

Pre-failure mechanical properties of each interface is derived from a homogenised elastic modulus that is endowed from each architecture~\cite{gibson_ashby_1997}. We consider a DCB configuration, where an ArchI is sandwiched between two beams. While beams are loaded perpendicularly to their plane, the interface intermediates load transfer and it can be regarded as being subjected to mode I loading conditions. We shall assume a simplification to this problem, so as to ensure the validity of the Winkler foundation approximation~\cite{dillard2018review}. Irrespective of the cell type, a single member of each geometry is identified as a truss of cross-section $A^*=\pi r^2$, where $r$ is the truss radius, and the Young's modulus of the base material is given by $E^*$. If all cells have a characteristic length $s$, we define the dimensionless aspect ratio $\tilde{l}\equiv r/s$ between the truss' smallest dimension and the cell size.

Generally speaking, the homogenised elastic modulus of each architecture can be written as:
\begin{equation}
\label{eq:GeneralStiff}
    \tilde{E}_{\mathrm{i}} = \frac{E_{\mathrm{i}}}{E^*}=C_{\mathrm{i}} \,\tilde{l}{\,}^{\alpha_{\mathrm{i}}},
\end{equation}
where $E_{\mathrm{i}}$ are effective Young's moduli of specific architecture and the index $\mathrm{i}$ is introduced in order to denote respective choices of architecture, \emph{i.e.} $\mathrm{i}\in\{\mathrm{p},\mathrm{t},\mathrm{h}\}$ in 2D for \emph{pillar}, \emph{triangular} and \emph{hexagonal}, and $\mathrm{i}\in\{\mathrm{p},\mathrm{o},\mathrm{k}\}$ in 3D for \emph{pillar}, \emph{octet} and \emph{Kelvin}. The powers $\alpha_{\mathrm{i}}$ are related to the topology of the architecture, meaning that $\alpha_{\mathrm{i}}=2$ for \emph{stretching dominated} cells and $\alpha_{\mathrm{i}}=4$ for \emph{bending dominated} cells. 
\rv{Notice that a different result is obtained when assuming 2D lattice material to be extended throughout the entire width. In this case, the power-law is given by $\alpha_{\mathrm{i}}=1$ and $3$ for stretching and bending-dominated structures, respectively \cite{gibson_ashby_1997}. In this work, we have chosen to limit the geometry to the width of the interface by a strut diameter ($2r$) instead, resulting in an increase of one for each power.}

For pillar interfaces, both 2D and 3D, given that the trusses are initially aligned with the externally applied loading, the forces are considered to be transmitted entirely along the pillar's longitudinal direction. The corresponding cell stress, $\sigma_{\mathrm{p}}=E_{\mathrm{p}}\epsilon_{\mathrm{p}}$, on the cross-section area $A_{\mathrm{p}}=s^2$ must equate to that of the single member, \emph{i.e.} $\epsilon_{\mathrm{p}} A_{\mathrm{p}} E_{\mathrm{p}} = \epsilon^* A^* E^*$, where the respective strains of the pillar unit and cell are $\epsilon_{\mathrm{p}}$ and $\epsilon^*$. It is, therefore, found that $C_{\mathrm{p}}=\pi$ and $\alpha_{\mathrm{p}}=2$. The triangular and octet truss cells are stretching dominated under Maxwell counting principle~\citep{calladine1978buckminster}, for which it has been shown that the stiffness scales with the aspect ratio with a power $\alpha_{\mathrm{i}}=2$~\citep{gibson_ashby_1997}. The normalised stiffness of the triangular interface is given by $C_{\mathrm{t}} = \pi (3+\sqrt{2})/7 $, whereas for the octet truss $C_{\mathrm{o}} = 2\sqrt{2}\pi $. This is derived from the stiffness matrix method~\citep{DESHPANDE_EffectiveProOfOctet}, which results in the effective interface property in the vertical direction $E_{\mathrm{o}} = \sqrt{2}\pi E^*r^2/l^2$, where $l=s/\sqrt{2}$ is the length of a single truss. Hexagonal and Kelvin interfaces are both bending dominated~\citep{calladine1978buckminster}, which makes their stiffness scale in $\tilde{l}$ with a power $\alpha_{\mathrm{i}}=4$~\citep{gibson_ashby_1997,ZHU_PropOfKelvinCell}. For such cases, deformations are assumed to only come from bending of the incline trusses. The normalised stiffness of the hexagonal cell is, therefore, given by eq.~(\ref{eq:GeneralStiff}) with $C_{\mathrm{h}}=32\sqrt{3}\pi$, and the Kelvin cell yields $C_{\mathrm{k}} = 1536\pi\cos{(\phi/2)}^5/[\sqrt{2}\sin{(\phi/2)}^2+\sin{(\phi/2)}]^2$, where the angle $\phi = 4\arctan{[(\sqrt{23}-2\sqrt{2})/5]}$ determines the shape of the Kelvin cell. 
%%%%%%%%%%%%%%%%%%%%%%%%%%%%%%%%
\subsection{Cells' failure conditions}
%%%%%%%%%%%%%%%%%%%%%%%%%%%%%%%%
To incorporate the possible failure mechanisms in \emph{stretching dominated} architectures, we consider the material's tensile strength, $\sigma_{\mathrm{f}}$, and the critical buckling force, $F_c$. This leads to the two strain based failure modes, in tension and compression:
\begin{subequations}
\label{eq:FailureI}
\begin{align}
\label{eq:FailureITS}
    \epsilon_{\mathrm{i}}^{(t)} &= J_{\mathrm{i}} \frac{\sigma_{\mathrm{f}}}{E^*}\\
\label{eq:FailureICS}
    \epsilon_{\mathrm{i}}^{(c)} &= -K_{\mathrm{i}} \pi^2 \tilde{l}^2
\end{align}
\end{subequations}
where $J_{\mathrm{i}}$ and $K_{\mathrm{i}}$ are constants depending on the cell geometry. For the pillar interface, the force is applied in the axial direction of the truss leading to $J_{\mathrm{p}}=1$.  Euler's critical load for the pillar structure fixed in both ends are given by $F_{\mathrm{c}} =4 \pi^2E^*I^*/h_{\mathrm{p}}^2$, where $h_{\mathrm{p}}=ns$ is the height of the interface leading to $K_{\mathrm{p}}=1/n^2$, and $I^*$ is the second moment of the pillar cross-section area. For the triangular interface $J_{\mathrm{t}}=\sqrt{2} \pi/C_{\mathrm{t}}$ and, from the buckling critical loading, we obtain $K_{\mathrm{t}}=J_{\mathrm{t}}/2$. Similarly, for the octet truss interface $J_{\mathrm{o}}=4\sqrt{2} \pi / C_{\mathrm{o}}$ and $K_{\mathrm{o}}=J_{\mathrm{o}}/2$. 

For the \emph{bending dominated} interfaces, the failure of the material is assumed to occur when the bending stress of the inclined truss reaches the failure stress of the material, under either tension or compression loading, leading to the following bounds:
\begin{subequations}
\label{eq:FailureIB}
\begin{align}
\label{eq:FailureITB}
    \epsilon_{\mathrm{i}}^{(t)} &= J_{\mathrm{i}} \frac{\sigma_{\mathrm{f}}}{E^*} \tilde{l}^{-1}\\
\label{eq:FailureICB}
    \epsilon_{\mathrm{i}}^{(c)} &= -\epsilon_{\mathrm{i}}^{(t)}
\end{align}
\end{subequations} 
where $J_{\mathrm{h}}=16\pi/(3C_{\mathrm{h}})$ and $J_{\mathrm{k}}=\sin{(\phi/2)}/ 12 [1-\sin{(\phi/2)}^2]$ for the hexagonal at the octet truss interface respectively. 

%%%%%%%%%%%%%%
\subsection{Analytical representation of DCB}
%%%%%%%%%%%%%%

% %%%%%%%%%%%%%%%%%%%%
% \begin{figure}[ht]
%     \centering
%     \includegraphics[height=7.5cm]{Figures/Beam_cases.png}
%     \caption{Unit cell, for the 2D figures the dimensions of the unit cells are $s\times2\times s$, for the 2D unit cells the dimensions is given on the figure. (a): Pillar 3D. (b): octet truss. (c): Kelvin cell (tetrakaidecahedron). (d) Pillar 2D. (e) Triangular structure.  (f) Hexagonal.}
%     \label{fig:Beam_cases}
% \end{figure}
% %%%%%%%%%%%%%%%%%%%%
We contrast results from two beam models, namely the Euler–Bernoulli beam theory (EBT) and the limit case of zero curvature beam, or a rigid. The focus is on the most important points needed for analysis, with additional details of calculations provided in \ref{Appendix_A}. The ArchIs are represented by the Winkler type elastic foundation, with homogenised materials properties of the cells (Fig. \ref{fig:Scope}) given by eq.~\eqref{eq:GeneralStiff}. The critical forces for failure of the interfaces are determined, together with the compliance of the DCB. Three Boundary Value Problems (BVP) are considered as outlined in Table \ref{tab:BC_for_beams}. Here, $Q_{xz}(x)$ and $M(x)$ are respectively the shear force and the moment in the beam, and $w(x)$ is the displacement of the beam in the $z$-direction. 
% \begin{figure}[h]
%     \centering
%     \includegraphics[width=0.9\textwidth]{Figures/Analytical/winklers_foundation.jpg}
%     \caption{Model of the DCB using symmetry. The interface material is represented by the Winkler foundation.}
%     \label{fig:setup_winklerfoundation}
% \end{figure}

\begin{table}[h]
\caption{BVP investigated in this study.}
\centering
\begin{tabular}{c|c|c|c}
\hline \hline
  \textbf{BVP}      & \textbf{Case A}   & \textbf{Case B}    & \textbf{Case C}                                                                                          
\\ \hline 
\textbf{Beam model}           & EBT     & EBT   & Rigid beam  
\\ \hline \hline
\textbf{BC} & 
\begin{tabular}[c]{@{}l@{}}$Q_{xz}(x=0)=F$ \\ $M(x=0) = 0$\\ $w'(x=L) = 0$\\ $w(x=L) = 0$\end{tabular} & \begin{tabular}[c]{@{}l@{}}$Q_{xz}(x=0)=F$ \\ $M(x=0) = 0$\\ $Q_{xz}(x=L) = 0$\\ $M(x=L) = 0$\end{tabular} & \begin{tabular}[c]{@{}l@{}}$Q_{xz}(x=0)=F$ \\ $M(x=0) = 0$\end{tabular} 
 \\ \hline \hline
\end{tabular}
\label{tab:BC_for_beams}
\end{table}
%%%%%
In Case A, an infinitely-long interface is assumed, i.e. $L\rightarrow\infty$, and $L\gg a_0$, where $L$ is the total length of the beam and $a_0$ is the initial crack length, such that there is neither rotation nor deflection of the beam far away from the point $x=0$, where the force $F$ is applied. 
In Case B, the length of the interface section is assumed finite and of length $L$. Hence, neither force nor moment can be transmitted at $x=L$. 
In Case C, the beam is behaving as infinitely rigid in both shear and bending, which is adequate for many practical cases including soft and/or short interface regions or crack/damage approaching the terminal beam edge, \emph{i.e.} $L-a\rightarrow0$, where $a$ is the current crack length.
Notice that, practically, the three BVP are likely to be mapped onto one another during the crack growth.

The beams shown in Fig.~\ref{fig:Scope} (a) are of rectangular cross section, with the second moment of area given by $I=bh^3/12$, where $b$ is the width and $h$ is the thickness of the beam. \rv{The material properties of these beams is given by their Young's modulus, which is here denoted by $E$.}
%$E$ and $E_{\mathrm{i}}$ are the elastic moduli of the beams and the interfaces, respectively. 
The total thickness of the interface is given by $h_{\mathrm{i}}$.
The elastic Winkler foundation should be regarded as distributed linear elastic springs with the stiffness $k_i=2E_{\mathrm{i}} b/h_{\mathrm{i}}$, given a distributed load on the beam of $q(x)=k_i w(x)$ for $x>a$ \citep{dillard2018review}. Assuming the Euler-Bernoulli beam kinematics, the governing equation is given by \citep{dillard2018review}: 
\begin{equation}
%EI\diff[4]{w}{x}= p(x)-q(x) \rightarrow
\diff[4]{w}{x} = - H(x-a) \frac{2 E_{\mathrm{i}} b}{h_{\mathrm{i}} EI}w(x),
    \label{eq:Governing_eq_EB}
\end{equation}
%where $p(x)=0$ is the applied load (force per unit length) and $q(x)$ is the reaction of the e.g. the Winkler foundation. 
where $H(x-a)$ is the Heaviside step function, which determines the position of the crack front. By defining $\lambda_{\mathrm{i}}\equiv\left({2 h_{\mathrm{i}} EI}/{E_{\mathrm{i}} b}\right)^{1/4}$, the governing equation can be written as
\begin{equation}
     \diff[4]{w}{x}+\frac{4}{\lambda_{\mathrm{i}}^4} w H(x-a) =0,
     \label{Eq:EB_governing_eq}
\end{equation}
where $\lambda_{\mathrm{i}}$ is a characteristic wavelength that determines the location for changes in load direction along the interface, from being loaded under tension to compression. Rewriting $\lambda_{\mathrm{i}}$ using the normalised parameters $\tilde{h}_{\mathrm{i}}\equiv h_{\mathrm{i}}/h$, $\tilde{E}_i \equiv E_{\mathrm{i}}/E^*$ and $\bar{E}\equiv E/E^*$ gives:
%--------------
\begin{equation}
\label{eq:Lambda_i}
    \lambda_{\mathrm{i}} = \lambda_0 \left( \frac{\tilde{h}_i}{\tilde{E_{\mathrm{i}}}/\bar{E}} \right) ^{1/4},
\end{equation}
%--------
where $\lambda_0 \equiv h/6^{1/4}$ is the characteristic wavelength for a DCB when the interface is of zero thickness~\citep{kanninen1973augmented}. 
Eq.~\eqref{Eq:EB_governing_eq} is solved separately for $x\leq a$ and $x>a$. The two solutions are then connected using continuity of the point of intersection $x=a$ (details provided in \ref{Appendix_A}).

On the other hand, for the rigid beam case, we may consider that the deflection is represented by a linear equation. This deflection is found from minimisation of the potential energy, $\Pi=U-W$. 
Since the beam is assumed to be rigid, the strain energy will be stored only in the ArchIs, represented by the Winkler foundation, thus leading to $U=1/2 \int_a^L [2 E_{\mathrm{i}} b w(x) /h_{\mathrm{i}}] dx$. The work done by the external forces is $W=F\delta$. Further details can be found in \ref{Appendix_A}.

%%%%%%%%%%%%%%%%%%%%%%%%%%%%%%%%%%%%%%%%%%%%%%%%%%%%%%%%%%%%%%% Critical force %%%%%%%%%%%%%%%%%%%%%%%%%%%%%%%%%%%%%%%%%%%%%%%%%
\subsection{Critical Force}
 %%%%%%%%%%%%%%%%%%%%%%%%%%%%%%%%%%%%%%%%%%%%%%%%%
The critical force, defined as the force at which failure of the interface begins, is determined by considering the critical strain of the interface, given eqs.~\eqref{eq:FailureI} for each choice of geometry. The maximum deflection of the interface, and the related strain, are positioned at $\tilde{x}=\tilde{a}$, where $\tilde{x}\equiv x/\lambda$ and $\tilde{a}\equiv a/\lambda$ are the normalised position and the crack length, respectively. Likewise, $\tilde{L}\equiv L/\lambda$ is the normalised length of the DCB, and the normalised deflection is given by $\tilde{w}=w/h_{\mathrm{i}}$. Thus, the normalised maximum deflection is found to be given by 
\begin{equation}
\label{eq:max}
\begin{aligned}
\tilde{w}_{\mathrm{max}} = \frac{\tilde{F}}{2}
\times
    \begin{cases}
        (1+\tilde{a}) 
        & \text{Case A}
        \\
        \frac{e^{4\tilde{L}}(\tilde{a}+1)
                            + e^{4\tilde{a}} (\tilde{a}-1) 
                            - 2e^{2\tilde{L}+2\tilde{a}}
                            [\tilde{a} \cos{(2\tilde{L}-2\tilde{a})} +\sin{(2\tilde{L}-2\tilde{a})]}} {e^{4\tilde{L}}+e^{4\tilde{a}}+2e^{2\tilde{L}+2\tilde{a}}
                            [\cos(2\tilde{L}-2\tilde{a})-2]} 
        & \text{Case B}
        \\                        \frac{2\tilde{L}^2-\tilde{a}^2-\tilde{L}\tilde{a}}{(\tilde{L}-\tilde{a})^3}
        & \text{Case C},
     \end{cases}
\end{aligned}
\end{equation}
%%%%%%%%%%%%%%% Minimum strain %%%%%%%%%%%%%%%%%%%%%%%
where the applied normalised force $\tilde{F}$ is defined as \cite{athanasiadis2021confined}:
\begin{equation}
    \Tilde{F}\equiv \left( \frac{\lambda_{\mathrm{i}}^3}{EIh_{\mathrm{i}}}\right)F.
\end{equation}
Such definition of $\tilde{F}$ is used later to normalise the obtained numerically results.
Depending on the choice of BVP, the minimum deflection of the interface region $(\tilde{x}>\tilde{a})$ can be found either within the interface region or at the end of the DCB specimen, $\tilde{x}=\tilde{L}$. 
For the Case A, the loading lead to a compression zone within the interface region for which the beam deflection is negative with the minimum deflection found by solving the equation ${d\tilde{w}}/{d \tilde{x}}=0$. 
Assuming a finite length of the interface section, as in Case B, the minimum deflection is localised either inside the interface region or at $\tilde{x}=\tilde{L}$, depending on the geometry and stiffness of the beam. In this situation, the normalised minimum deflection $\tilde{w}_{\mathrm{min}}$ is determined numerically.

The normalised compliance, using minimum deflection, is defined as $\tilde{c}_{\mathrm{min}}\equiv\tilde{w}_{\mathrm{min}}/\tilde{F}$. This will later be used to determine the critical force. Considering Case C, the minimum deflection will be localised at $\tilde{x}=\tilde{L}$. Therefore, its value can be found analytically: 
\begin{equation}
\label{eq:min}
 \tilde{w}_{\mathrm{min}} = -\frac{\tilde{F}}{2} \times
    \begin{cases}
    \sqrt{\frac{1}{2}+\tilde{a}(1+\tilde{a})}e^{\arctan(1+2\tilde{a})-\pi}
    & \text{Case A}
    \\
    \frac{\tilde{L}^2-2 \tilde{a}^2+\tilde{L}\tilde{a}}{(\tilde{L}-\tilde{a})^3}
    & \text{Case C.}
    \end{cases}
\end{equation}

%%%%%%%%%%%%%%% Failure %%%%%%%%%%%%%%%%%%%%%%%%%%%%%%%
From the expressions derived for the maximum, eq.~\eqref{eq:max}, and minimum, eq.~\eqref{eq:min}, deflections together with the critical strain in both tension and compression, eqs.~\eqref{eq:FailureITS}--\eqref{eq:FailureICB}, the critical force is found from $\epsilon_{\mathrm{i}}^{(t)}=2 w_{\mathrm{max}}$ and $\epsilon_{\mathrm{i}}^ {(c)}=2 w_{\mathrm{min}}$. Therefore, the critical force under tensile loading of the interface is given by
\begin{equation}
    \tilde{F}_c^{(t)} = \epsilon_{\mathrm{i}}^{(t)} \times
    \begin{cases}
                \frac{1}{1+\tilde{a}} & \text{Case A}
        \\
        \frac{e^{4\tilde{L}}+e^{4\tilde{a}}+2e^{2\tilde{L}+2\tilde{a}}
        [\cos(2\tilde{L}-2\tilde{a})-2]}{e^{4\tilde{L}}(\tilde{a}+1)
                            + e^{4\tilde{a}} (\tilde{a}-1) 
                            - 2e^{2\tilde{L}+2\tilde{a}}
                            [\tilde{a} \cos{(2\tilde{L}-2\tilde{a})} +\sin{(2\tilde{L}-2\tilde{a})]}}
        & \text{Case B}
        \\
        \frac{(\tilde{L}-\tilde{a})^3}{2\tilde{L}^2-\tilde{a}^2-\tilde{L}\tilde{a}}
        & \text{Case C},
    \end{cases}
    \label{Eq:Fctilde_crit_t}
\end{equation}
whereas the critical force for failure of the interface region under compression loading of the interface is 
\begin{equation}
 \tilde{F}_c ^{(c)}= -\epsilon_{\mathrm{i}}^ {(c)}
    \begin{cases}
     \frac{e^{\pi - \arctan(1+2\tilde{a})}}{\sqrt{\frac{1}{2}+\tilde{a}(1+\tilde{a})}}
    & \text{Case A}
    \\
    -\frac{1}{2 \tilde{c}_{\mathrm{min}}}
    & \text{Case B}
    \\
    \frac{(\tilde{L}-\tilde{a})^3}{-\tilde{L}^2+2 \tilde{a}^2-\tilde{L}\tilde{a}}
    & \text{Case C.}
    \end{cases}
    \label{Eq:Fctilde_crit_c}
\end{equation}
%%%%%%%%%%%%%%%%%%%%%%%%%
where $\lambda_{\mathrm{i}}$ is calculated from eq. (\ref{eq:Lambda_i}).
%%%%%%%%%%%%%%%%%%%%%%%%%%%%%%%%%%%%%%%%%%
\subsection{Pre-failure compliance}
%%%%%%%%%%%%%%%%%%%%%%%%%%%%%%%%%%%%%%%%%%
The compliance for the DCB is defined by $c\equiv \delta/F$, where $F$ is the applied force and $\delta$ is the displacement of the left end on the bean shown in Fig. \ref{fig:Scope}. The normalised compliance is found by using the normalised force, $\tilde{F}$, and displacement, $\tilde{\delta}=\delta/h_{\mathrm{i}}$ and, hence, $\tilde{c} \equiv \tilde{\delta}/\tilde{F}=cEI/\lambda^3$. Then, considering e.g. Case A: 
\begin{equation}
\tilde{c} =
     \frac{1}{6}\left[1+2(1+\tilde{a})^3\right].
     \label{eq:normalized_compliance}
\end{equation}
%%%%%%%%%%%%%%%%%
\section{Numerical model}
%%%%%%%%%%%%%%%%%%%%%%%%%%%%%%%%%%%%
\subsection{Numerical implementation}
%%%%%%%%%%%%%%%%%%%%%%%%%%%%
The main scope of the present work is numerical modelling of damage of the interface with metamaterial cells presented in Fig. \ref{fig:Scope}, and thus complementing findings of \citep{athanasiadis2021confined,The_toughness_of_mechanical_metamaterials} with non-existing damage growth predictions. For that, the finite element method (FEM) is used.
\rv{The adherent thicknesses $h=h_{\mathrm{i}}$ is used  with the interface height set to $h_{\mathrm{i}} = \SI{20}{mm}$ and the initial crack length $a_0=10h$}.
The pre-processing, solving and post-processing were carried out in COMSOL Multiphysics (v5.8) for both the 2D and 3D geometries. 
For the 2D case, the Euler-Bernoulli beam elements are used for modelling both the adherents and the cell trusses. 
For the 3D case, the Euler-Bernoulli beam elements are used only for the cell trusses, while the adherents are modelled via the shell elements.
A linear-elastic material model is used for the adherents assuming the Young's modulus $E=\SI{100}{GPa}$ and Poisson's ratio $\nu=0.3$---the values corresponding roughly to titanium and its alloys or carbon fibre composites. The initial modulus of the material of the interface $E^*=\SI{5}{GPa}$, while the Poisson’s ratio is taken as $\nu^*=0.3$---roughly properties of structural polymers. Note, that the absolute values are here provided only for reproducibility reason and out of numerical implantation necessity. Our theoretical framework \cite{athanasiadis2021confined} removes such limitation allowing comparison of any geometries and materials as long as kinematic assumptions are obeyed. 
%%%%%%%%%%%%%%%%%%%%%%%%%%%%
\subsection{Damage model}
%%%%%%%%%%%%%%%%%%%%%%%%%%%%
Modelling material failure without impeding progressing damage zone is numerically challenging and computationally expansive. Such costs are even higher in the case of architected materials with complex cells geometries and multiple trusses per cell. Practically, once one of the cell trusses fails its stiffness can be considered as being zero. However, this provokes numerical instability arising from singularities in the numerical solution. To avoid such situation, in the current model the stiffness of the material is reduced $k=10^6$ times upon failure, thus, making the material sufficiently `soft'. Note, that constant $k$ will depend on the number of beam elements/trusses $n$ used within the model. In the present case $n$ varies for different cells but always fulfil $n\ll k$, and, thus, the accumulated stiffness effect of failed trusses is negligible. As a conservative ``rule-of-thumb'', we suggest keeping the ratio $k/n\geq 100$. The material model for the interface material becomes:
%%%%%%%%%%%%%%%%%%%%%%%%%%%%
\begin{equation}
  E(\epsilon)=\begin{cases}
    E^*/k,  & \text{if } \epsilon<\epsilon_{\mathrm{f}}^{(c)},\\
    E^*{\,\,\,\,\,\,},            & \text{if } \epsilon_{\mathrm{f}}^{(c)} \leq \epsilon \leq \epsilon_{\mathrm{f}}^{(t)},\\
    E^*/k,  & \text{if } \epsilon>\epsilon_{\mathrm{f}}^{(t)}
  \end{cases}
\end{equation}
%%%%%%%%%%%%%%%%%%%%%%%%%%%%
resembling damage model formulations. Here, $\epsilon$ is the axial strain of the beam elements, $\epsilon_{\mathrm{f}}^{(t)}$ and $\epsilon_{\mathrm{f}}^{(c)}$ are the failure strain under tension and compression, respectively.
For stretching dominated cells, the failure under tension loading occurs when the ultimate tensile stress $\sigma_{\mathrm{f}}$ is reached, $\epsilon_{\mathrm{f}}^{(t)}={\sigma_{\mathrm{f}}}/{E^*}$. Since the truss is considered slender, the failure under compression loading is expected when the buckling load $F_{\mathrm{f}}=({E^*\pi^3r^4})/{s^2}$ is reached, \emph{i.e.} $\epsilon_{\mathrm{f}}^{(c)}={F_{\mathrm{f}}}/({E^* \pi r^2})=-\pi\Tilde{l}^2$. For the bending dominated cells, the truss fails due to material failure under both tension and compression, hence $\epsilon_{\mathrm{f}}^{(t)}=-\epsilon_{\mathrm{f}}^{(c)}={\sigma_{\mathrm{f}}}/{E^*}$ including stresses due to bending.
%%%%%%%%%%%%%%%%%%%%%%%%%%%%
\subsection{Boundary conditions}
%%%%%%%%%%%%%%%%%%%%%%%%%%%%
All the simulations are performed under assumption of centre line symmetry. For the 2D case, following Fig. \ref{fig:Scope}, the symmetry around the $xy$-plane is considered, while for the 3D case both the $xy$- and $xz$- symmetry planes are used. 
When using the symmetry plane, some of the beams are cut into half. These are modelled as semicircles with the second moment of inertia with respect to the axes $x'$ and $y'$, given by $I_{x'}=I_{y'}=\frac{\pi r^4}{8}$, where $r$ is the radius of the truss.

The displacement $\delta$ is applied on the left side of the beam following Fig. \ref{fig:Scope}. The calculation is then carried by increasing $\delta$ and includes geometrical non-linearity. Two non-linear solvers are used \emph{Constant Newton} (with a damping factor of 1 which was found adequate for majority of cases) or \emph{Automatic Newton} (computationally more expensive as the the damping factor is reevaluated at each step. In the present case it varied  in the range of $0.1-1$) depending on the solution convergence.
%%%%%%%%%%%%%%%%%%
\section{Results and Discussion}
%%%%%%%%%%%%%%%%%%
\subsection{Compliance and beam kinematics prior to failure}
\begin{figure}[ht]
    \centering
    \includegraphics[width=7cm]{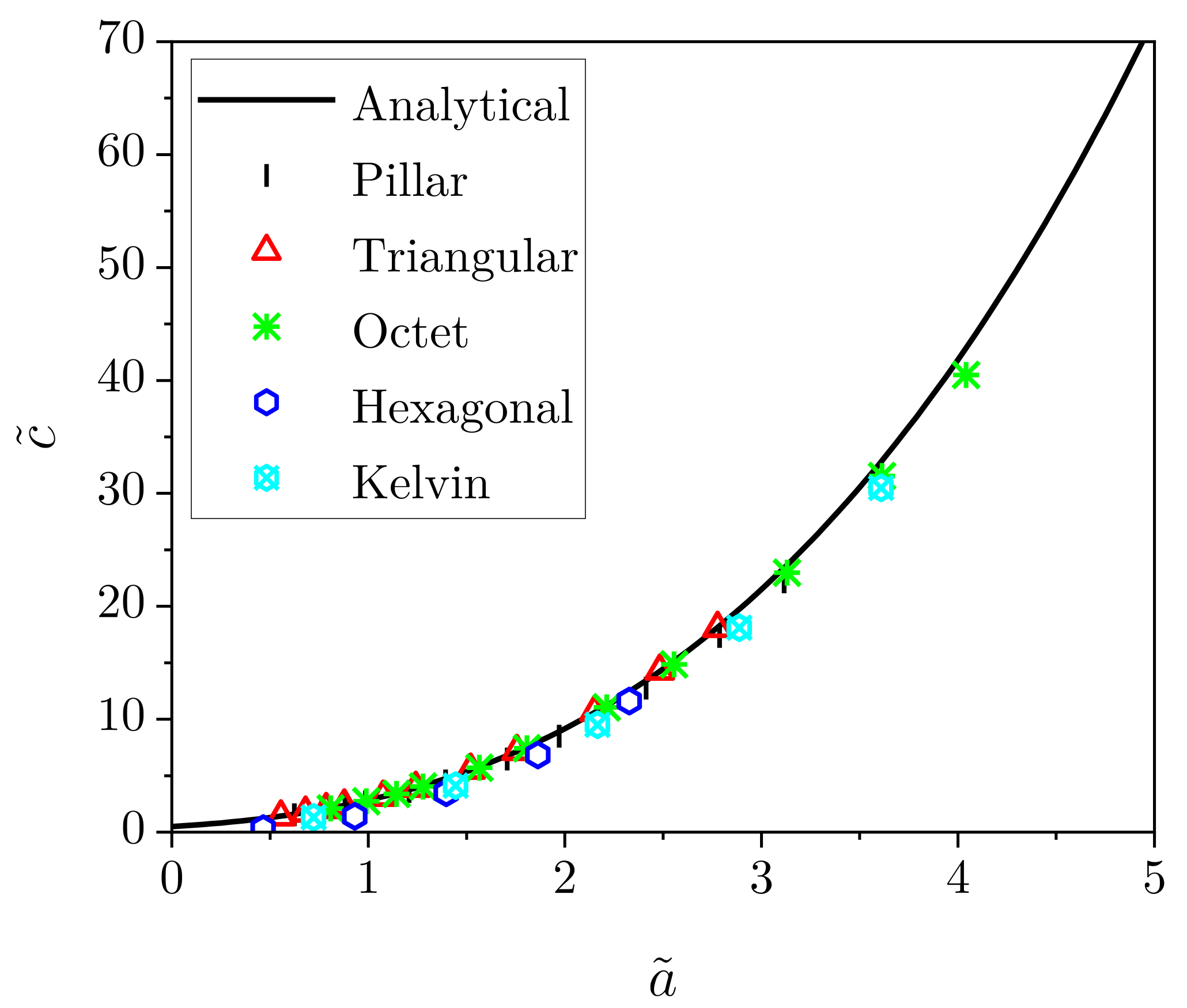}
    \caption{The initial normalized compliance for the DCB using the different interfaces geometries. Linear scale of the axis. The markers show the FEM numerical results and the solid-line show the analytical result from eq.~\eqref{eq:normalized_compliance}.}
    \label{fig:Compliance}
\end{figure}

%%%%%%%%%%%%%%%%%%
To verify the homogenisation steps of the proposed model, Fig. \ref{fig:Compliance} gathers the elastic loading compliance data for all the investigated cells and compares the numerical data with the analytical result of eq. (\ref{eq:normalized_compliance}). All the data collected here are for the DCB specimen with long interface sections (only Case A, Table \ref{tab:BC_for_beams}, is considered). 
%%%%%%%%%%%%%%%%%%%%
%%%%%%%%%%%%%%%%%%%%

Fig. \ref{fig:Compliance} shows a very good agreement between the analytical and the numerical model is obtained. However, some deviation can be expected for the case of extremely short cracks and bending dominated Kelvin and hexagonal cells. For short cracks the beam bending is very limited. However, the exerted rotation breaks the initial symmetry of the cells. For the bending cells such situation leads to conversion from bending to axial stress, and as a result lower compliance/higher stiffness results from the FEM. Such local/topological effects are not mirrored by the homogenised model and the Winkler formulation. The homogenisation assumes that the symmetry of the cell is maintained, thus, all the trusses undergoes only bending loading mapped to effective foundation stiffness---this is not the case one the symmetry is broken. In addition, within Winkler foundation formulation the stress is not transferred between the cells---the broken symmetry implies additional interaction beyond the limit of the model used. Similar effect on compliance is not found for all the cells experiencing tensile stress state.

\begin{figure}[ht]
    \centering
    \includegraphics[width=12.5cm]{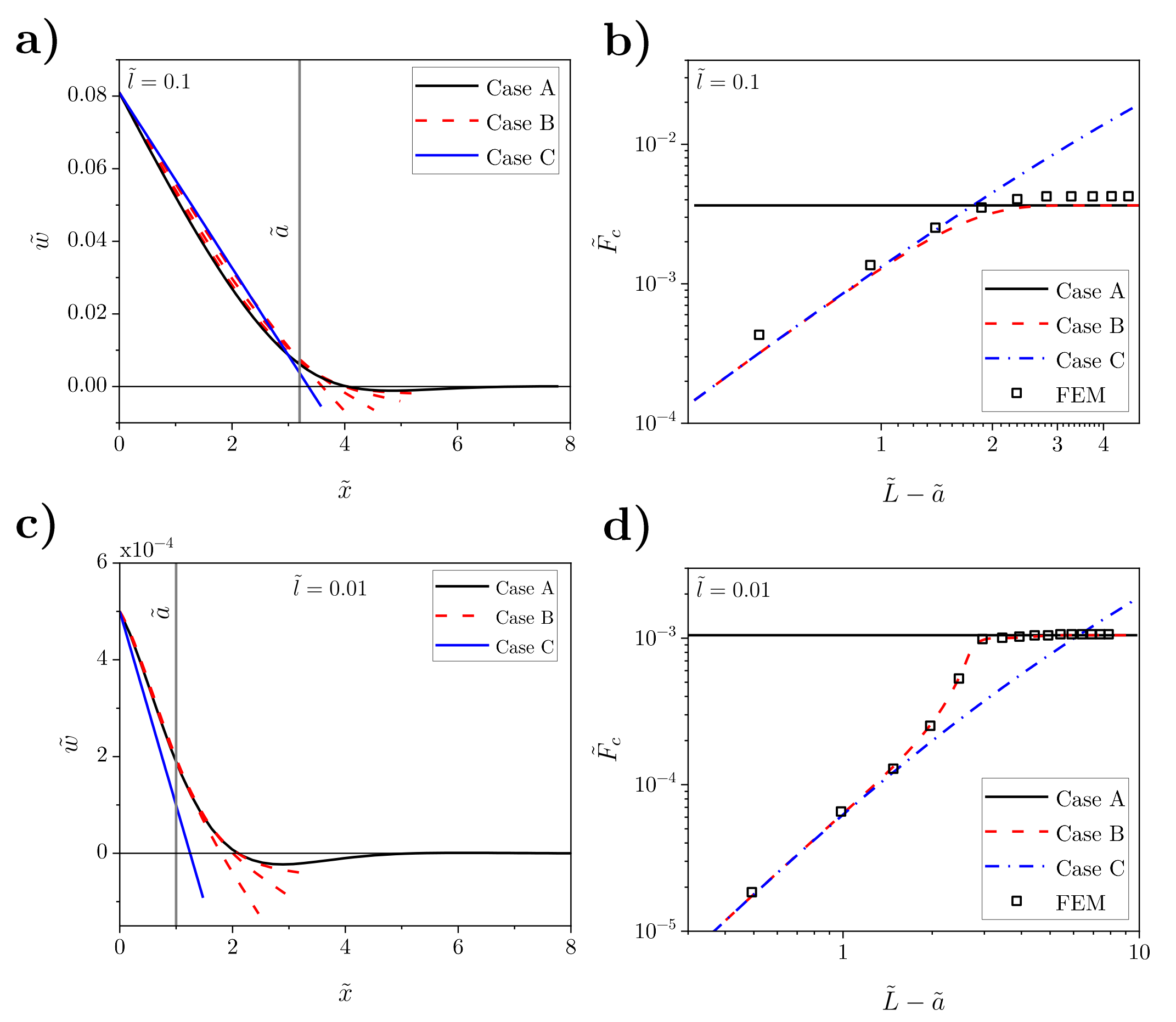}
    \caption{Deflection of the beam along the beam (a) and (c) for two different aspect ratios $\tilde{l}$ and BVPs (b) and (d) Respective fracture forces. The pillar interface is here used for the FEM, where for $\tilde{l}=0.1$ the interface is failing under tensile loading when $\tilde{l}=0.01$, the interface is failing under compression lading.}
    \label{fig:Deflection}
\end{figure}

In Fig. \ref{fig:Deflection} deflection along the beam, (a) and (c), and fracture forces, (b) and (d), for two different aspect ratios $\tilde{l}$ and for the different BVP formulations investigated from Table \ref{tab:BC_for_beams} are presented.

%%%%%%%%%%%%%%%%%%%%
%%%%%%%%%%%%%%%%%%%%

In Fig. \ref{fig:Deflection} (a) and (c) the kinematic behaviour of the beam, and beam--interface, is depicted outlining the effect of finite length of the interface region. Namely, given the known bending stiffness of the beam, the analytical problem may fall into different BVP. More specifically, Case A is valid for sufficiently long interface regions, ensuring the applicability of far field assumptions. If the interface region is of the order of characteristic length scale of the problem $\lambda_p$ BVP is changed. On the other hand, if the interface region is short, or sufficiently soft (high $\lambda_p$ value), the deflection is correctly captured only by BVP Case B, and, finally, C. It is important to notice that, for the given aspect ratio and material choice, the analytical predictions remain in excellent agreement with the numerical model. This is also the case for very small aspect ratios of the lattice truss, Fig. \ref{fig:Deflection} (c). Figs. \ref{fig:Deflection} (b) and (d) shows how important it is to formulate the BVP appropriately. Given reduction of the interface region length $\tilde{L}-\tilde{a}$, the critical force undergoes a smooth transition between different formulations, with Case B capturing such effects. The importance stems also from the fact that, during damage propagation, the BVP will indeed change. By using the far-field formulation, Case A can lead to a significant overestimation of fracture onset forces and non-conservative design. Finally, the Case C formulation, rigid body motion of the beam, offers high accuracy for small values of $\tilde{L}-\tilde{a}$. This is useful information enabling design of experiments targeting respective BVP for which the data reduction scheme is straightforward.
%%%%%%%%%%%%%%%%%%%%
\subsection{Failure onset and damage propagation}
\begin{figure}[ht]
    \centering
    \includegraphics[width=12.5cm]{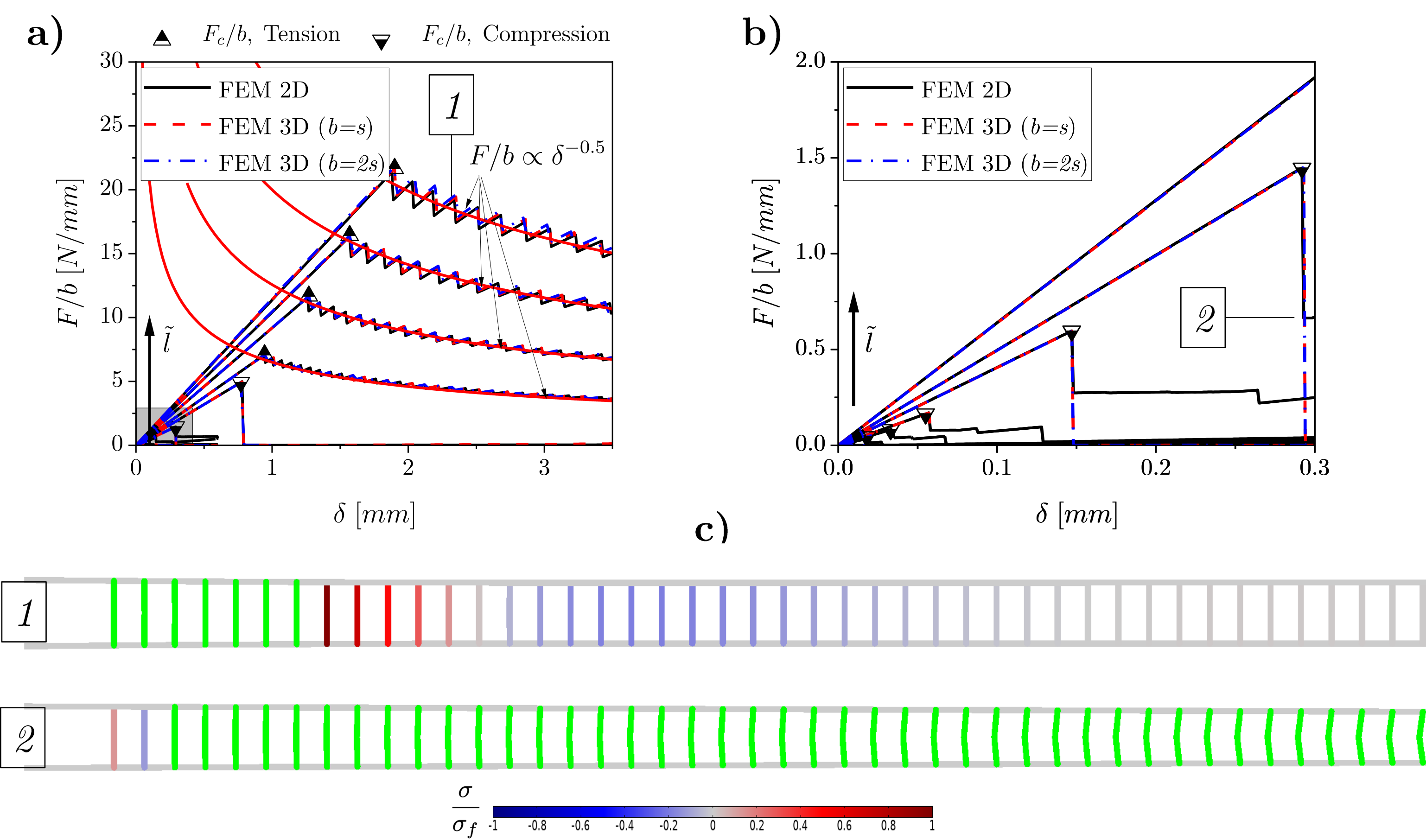}
    \caption{Failure of the pillar interface using both 2D and 3D models. (a) Force-displacement plot. (b) Zoom force-displacement plot showing failure of low aspect ratios interfaces. (c) Interface the two different aspects ratios at stages (1) and (2) corresponding to the force-displacement curves. Green colour denotes damaged trusses. The colour scale bar gives the normalised stress (by $\sigma_{\mathrm{f}}$) from compression (blue) to tension (red).}
    \label{fig:Pillar}
\end{figure}

%%%%%%%%%%%%%%%%%%%%
\rv{From this point forward we focus our attention to Case A, where an infinitely-long interface is assumed ($L\rightarrow\infty$) and $L \gg a_0$ in a DCB. This lies in the desire to focus on the critical region around the crack tip. By assuming an infinitely long interface, complexities associated with the beam's length are neglected. The condition $L \gg a_0$ ensures that far from the crack, the beam behaves as if there were no crack present, simplifying the overall analysis. We consider the microstructure to be perfect, ensuring that the symmetry conditions initially adopted remain valid during the fracture process. This approach is chosen to isolate and fully understand the fundamental mechanics that govern crack propagation, highlighting the influence of geometry, material properties, and applied forces in the immediate vicinity of the tip of the crack.}
Let us now look at the process with the most stable propagation of damage, which is achieved with the pillar interface, as seen in Fig. \ref{fig:Pillar}. In Fig. \ref{fig:Pillar} (a) the force-displacement curves for this case, with pillars of different aspect ratios---for both 2D and 3D geometries, where the BVP Case A is considered, are shown.
%%%%%%%%%%%%%%%%%%%%
%%%%%%%%%%%%%%%%%%%
For all the models a constant distance between the pillars was maintained as well as interface thickness. The 3D models are considered for $b=s$ and $b=2s$, giving one and two rows of pillars along the interface respectively. The critical force is marked by either upwards triangle or downwards triangle, each case indicating whether the failure is due to tension or compression, respectively. 
A very good agreement between the 2D and 3D models is evident for all the different aspect ratios $\tilde{l}$ depicted. For the pillar interfaces the force-displacement curve can be interpreted via two stages: (i) initial, linear, loading stage corresponding to the elastic response of material that upon reaching the critical force bifurcates to (ii) the non-linear, damage propagation, stage, governed by a power-law given by $F/b\propto\delta^{-0.5}$. 
Fig. \ref{fig:Pillar} (b) provides further details of propagation stage driven by compressive failure. In Fig. \ref{fig:Pillar} (c), the green colour is used to denote damaged pillars, while the colour gradient from red to blue denotes the tension-compression stresses within the interface region. 
For the cases with higher aspect ratio a stick-slip softening is observed during the damage propagation. It is noteworthy to remark that, upon externally applied loading, the failure features of ArchIs at the cell level do not necessarily related to fracture mechanics in the standard sense---\emph{i.e.} the individual trusses can fail in multiple ways corresponding to very different failure criteria. However, once integrated into the structure and constrained by other materials, the entire interface can be regarded as a ``crack'' growth plan. An otherwise continuous fracture process, for constant fracture energy, that is expected for homogeneous materials, is instead replaced by an iterative pillar failure process, with the onset and arrest forces oscillating around the mentioned fracture scaling relation given by $F/b\propto\delta^{-0.5}$. The magnitude and wavelength of this oscillation are directly related to the geometry of the cell. In the pillar case to $\tilde{l}$.  
On the other hand, for lower aspect ratios, instead of the growth we observe sudden snap-down phenomena. In these cases, we see a large section of the interface region fail due to the buckling phenomena.
%%%%%%%%%%%%%%%%%%%%
\subsection{Toughening mechanisms of complex interfaces}
\begin{figure}[ht]
    \centering
    \includegraphics[width=12.5cm]{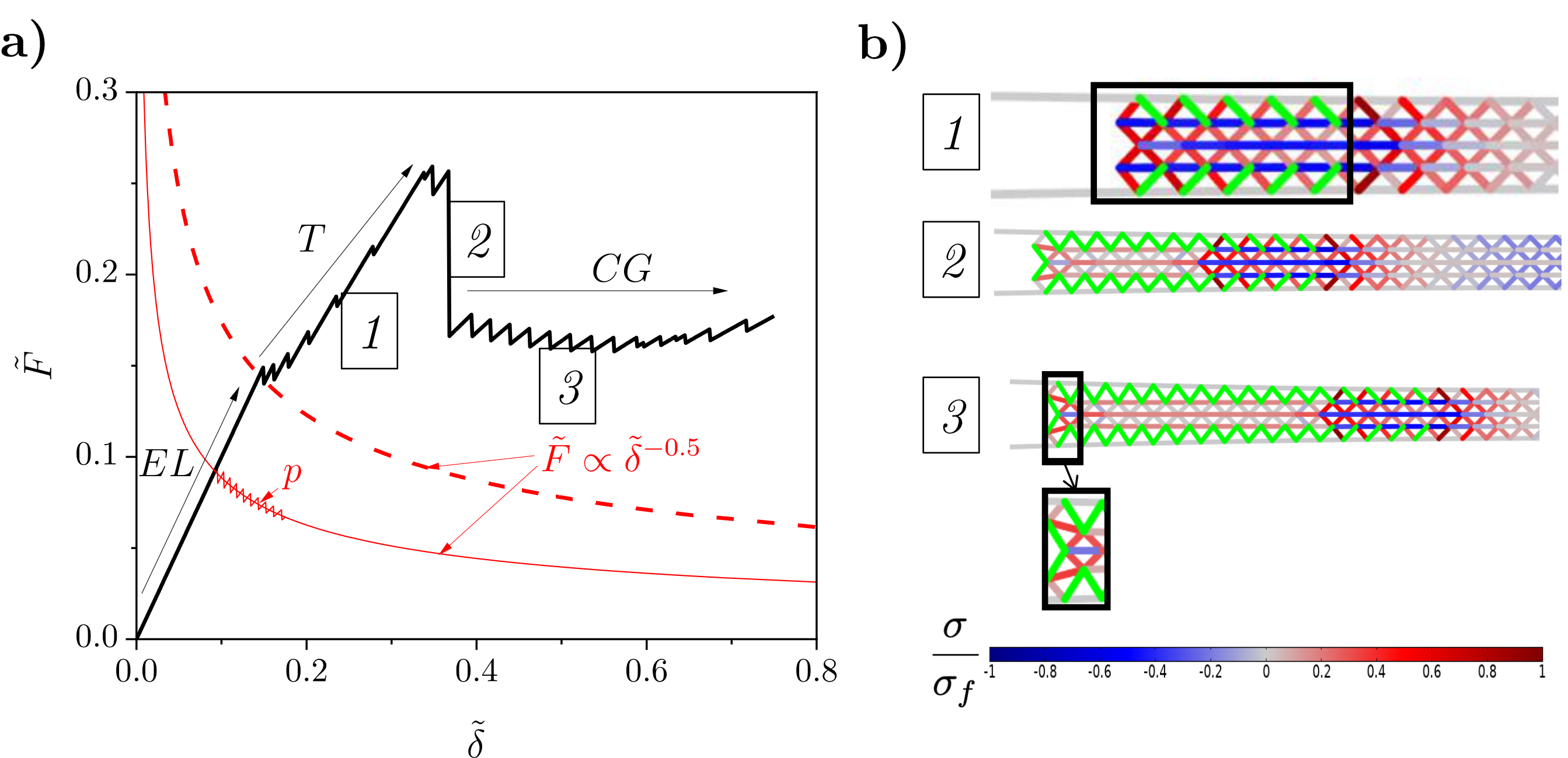}
    \caption{(a) Force-displacement curve triangular interface (blue) where $\tilde{l}=0.1$. The curve is divided into three sections; an elastic zone (EL), a toughening zone (T) and a zone for the crack growth (CG). The curve for the pillar interface (p) is included for comparison. (b) Stresses and damage inside the triangular interface. Green colour denotes damaged trusses. The colour scale bar gives the normalised stress (by $\sigma_{\mathrm{f}}$) from compression (blue) to tension (red).}
    \label{fig:Triangular}
\end{figure}

\begin{figure}[ht]
    \centering
    \includegraphics[width=12.5cm]{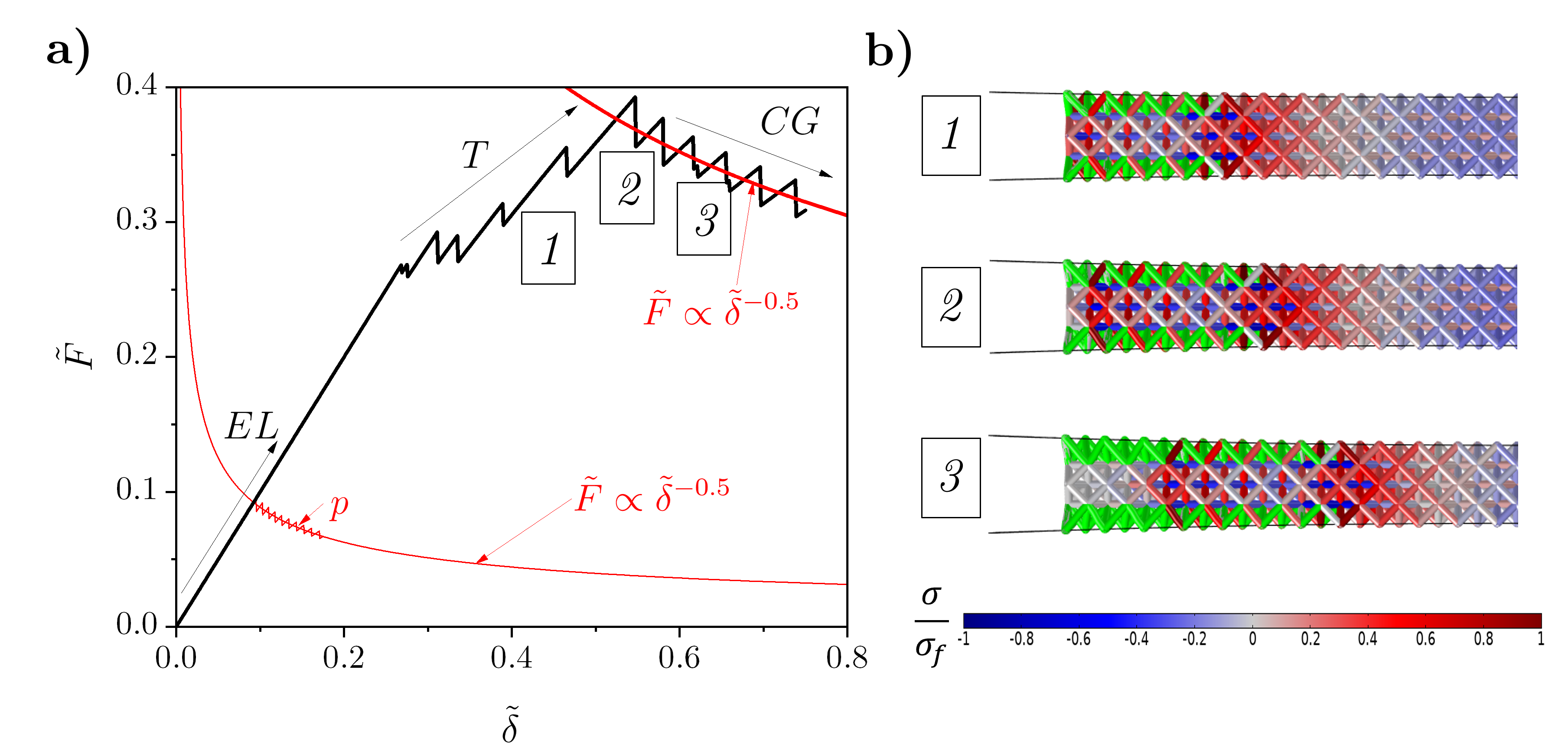}
    \caption{(a) Force-displacement curve of the octet interface (orange) where $\tilde{l}=0.1$. (b) Stresses and damage inside the octet interface. Green colour denotes damaged trusses. The color scale bar gives the normalised stress (by $\sigma_{\mathrm{f}}$) from compression (blue) to tension (red).}
    \label{fig:Octet}
\end{figure}

\begin{figure}[ht]
    \centering
    \includegraphics[width=12.5cm]{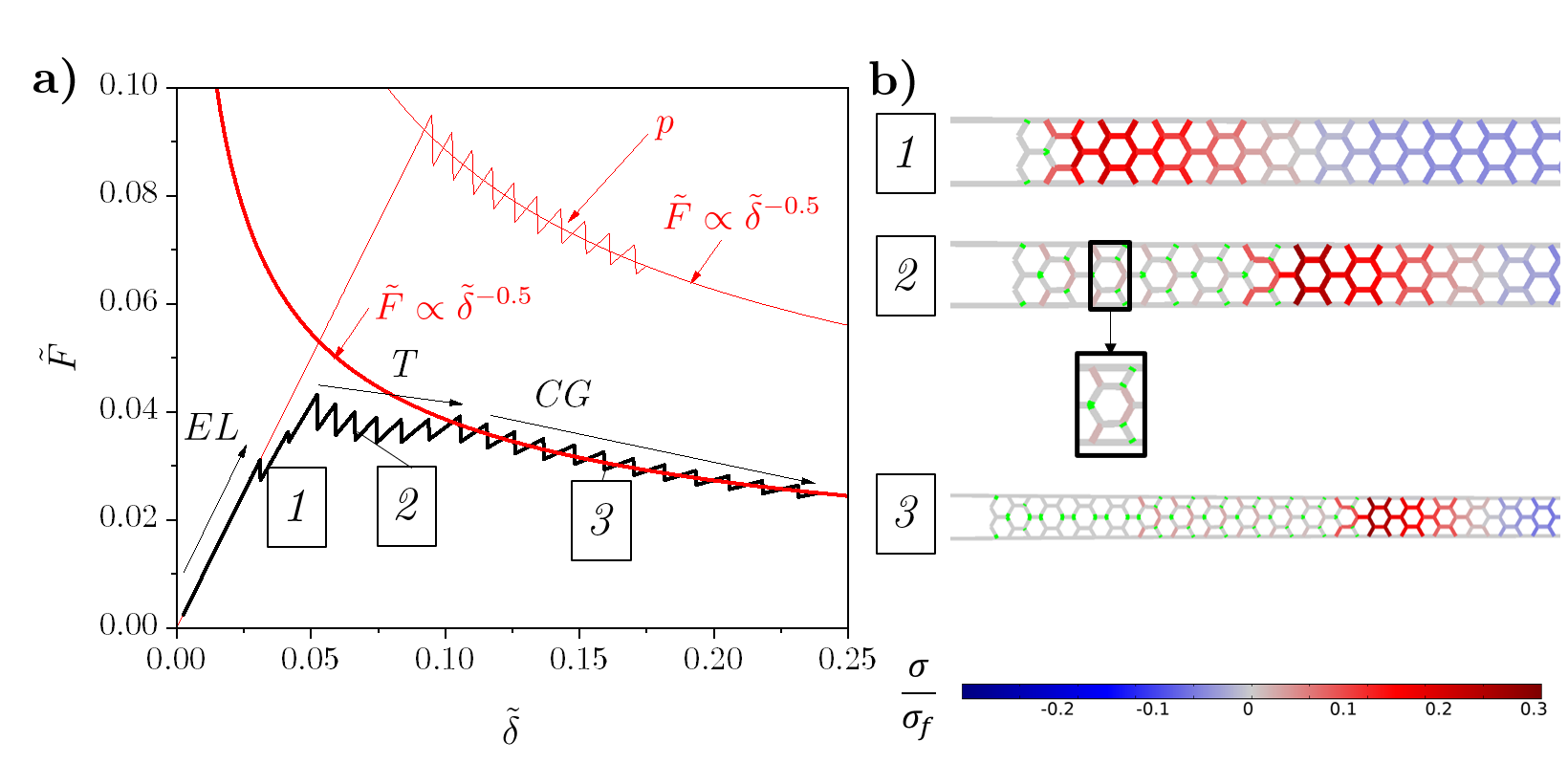}
    \caption{(a) Force-displacement curve of the hexagonal interface where $\tilde{l}=0.1$. The curve is divided into two sections; an elastic zone (EL) and a toughening zone (T+G). For comparison a pillar interface case is also plotted and denoted with p. (b) Stresses and damage of hexagonal interface at three different stages corresponding to (a). Green colour denotes damaged trusses. The colour scale bar gives the normalised axial stress (by $\sigma_{\mathrm{f}}$) from compression (blue) to tension (red).}
    \label{fig:Hexagonal}
\end{figure}

\begin{figure}[ht]
    \centering
    \includegraphics[width=12.5cm]{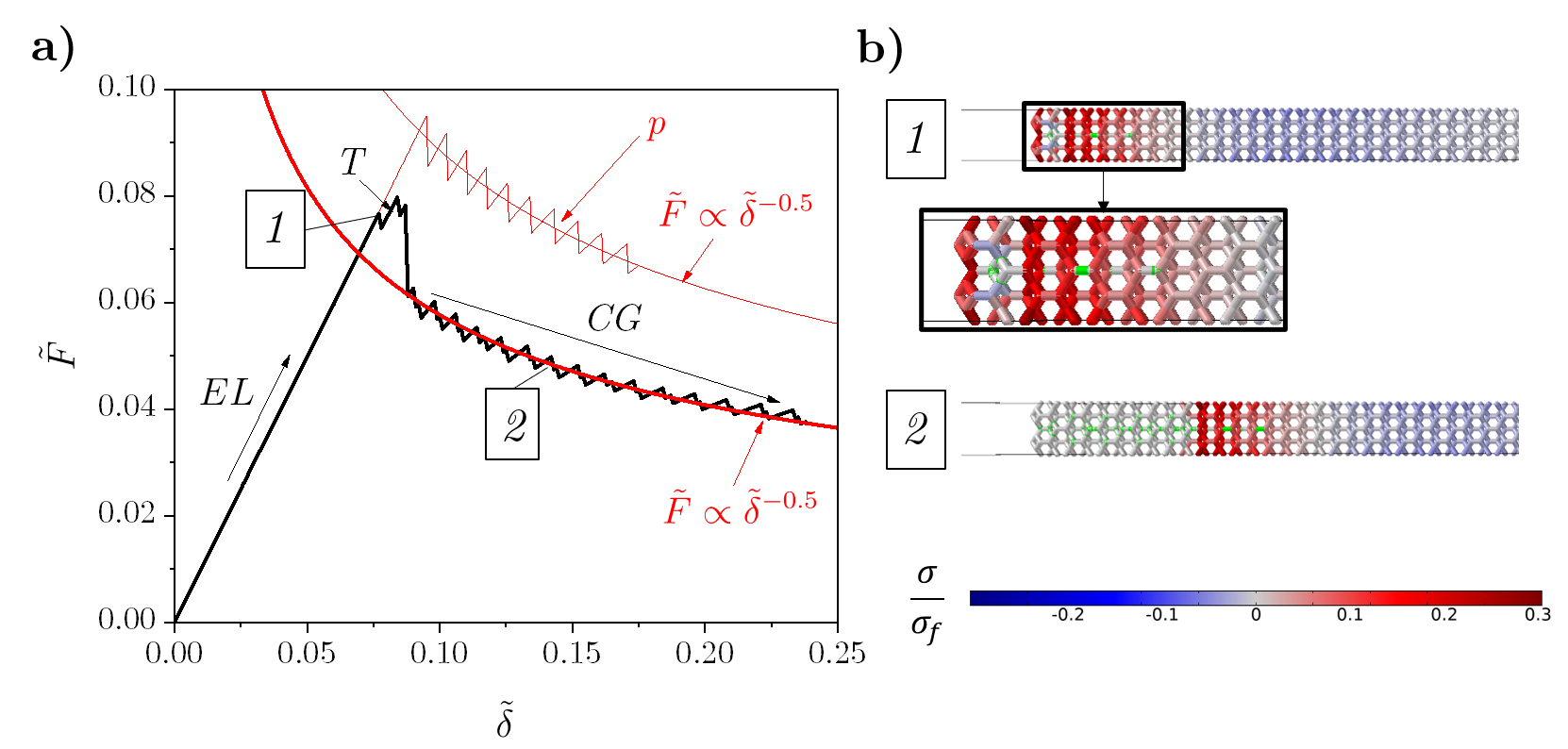}
    \caption{(a) Force-displacement curve of the Kelvin interface (purple) where $\tilde{l}=0.1$. The curve is divided into three sections; an elastic zone (EL), a toughening zone (T) and a zone for the crack growth (CG). The curve for the remaining curves are included for comparison. (b) Stresses and damage of Kelvin cell interface. Green colour denotes damaged trusses. The colour scale bar gives the normalised axial stress (by $\sigma_{\mathrm{f}}$) from compression (blue) to tension (red).}
    \label{fig:Kelvin}
\end{figure}

%%%%%%%%%%%%%%%%%%%%
We consider the damage propagation for the remaining, topologically more complex interfaces: triangular, octet truss, hexagonal and Kelvin. As before, we focus on the same BVP given by case A from Table~\ref{tab:BC_for_beams}. In Figs. \ref{fig:Triangular} to \ref{fig:Kelvin} the force-displacement curves for the mentioned interface types (with an aspect ratio $\tilde{l}=0.1$) are depicted. For better comparison, the force has been normalised the following way: $\tilde{F}\equiv F c_0/h_{\mathrm{i}}$, where $c_0$ is the initial compliance.
%%%%%%%%%%%%%%%%%%%%
%%%%%%%%%%%%%%%%%%%

Starting with Fig. \ref{fig:Triangular}, the stresses and damage stages of triangular interface are shown. For comparison, a corresponding $\tilde{l}$ curve for pillar interface is shown (denoted with letter `p'). As it was for the pillar interface, Fig. \ref{fig:Pillar}, the force increase linearly until failure of the first truss. Then, a bridging zone connecting the bottom and top plates is created, similar to the well-known from composite materials phenomenon. This allows the force to increase beyond the initial failure force, thus resulting in a toughening effect. The force continues to increase until a sudden drop, marking failure of some of the created bridging links. Notice that at this stage, for the triangular cells, the failure force is almost doubled, compared to the first truss failure loading, and that the stored energy would have quadrupled. The damage begins to propagate with characteristic stress pattern within the ArchI, visible in the second and third inset from the top of Fig. \ref{fig:Triangular} (b). By comparing the triangular case with the pillar interface, the $F\propto\delta^{-0.5}$ law is not followed. This leads us to conclude that the process is not self-similar and it can be split into the following physical phenomena: (i) linear loading below the initial failure load (denoted as EL); (ii) unfolding of topologically preferable links bridging the face/beam materials, thus resulting in significant toughening effects (T); (iii) failure of some of the bridges and creation of quasi self-similar damage zone composed of both the remaining bridges and leftover interface core; (iv) and finally crack growth (CG) of such a pattern. As these are novel observations, some additional comments should be added. Within the loaded portion of the interface, and especially at the very edge of the interface, the loading direction is skewed locally in relation to initial symmetry planes of the cell. Different features and damage patterns are to be expected depending on the side of the cell, \emph{i.e.} one side is free from direct interactions with other cells and thus more compliant and capable to higher extensions, while the other is constrained by such interactions. This effect is captured in Fig. \ref{fig:Triangular} (b) through the inset zooming the edge of the interface. Even if the main stress field, and its characteristic pattern, propagated by a distance of few interface thicknesses, the edge link is still carrying the load causing deviation from the fracture power-law. The significant effect is stemming from the fact that this link is unfolding and approaching pure tension conditions.

Similar behaviour is observed for the octet interface, as shown in Fig. \ref{fig:Octet}.
%%%%%%%%%%%%%%%%%%%
%%%%%%%%%%%%%%%%%%%
Here, the interface also experiences a toughening effect, while a bridging zone is created and, qualitatively, the same features of the load response curve could be recognised. However, by contrast with the 2D triangular interface, it is noticed that the CG phase presents a stable and self-similar process with the characteristic power low. The pillar cell interface is left for comparison and it is denoted by the letter p.

Figs. \ref{fig:Hexagonal} and \ref{fig:Kelvin} show the load response of the hexagonal and Kelvin interfaces, respectively.
%%%%%%%%%%%%%%%%%%%
%%%%%%%%%%%%%%%%%%%
%
As for the previous interfaces, both tension and compression zones could be found along the interface. As previously, loading the DCB with the hexagonal interface, Fig. \ref{fig:Hexagonal}, the force is initially increased in linear way until the point where the first truss fails. After this point, a bridging zone is created while the displacement $\tilde{\delta}$ is increased, which causes the creation of a toughening effect. However, neither steep increase nor sudden drop of the force are observed. Focusing on the remaining trusses in the bridging zone, it is shown that these trusses are creating a pattern, for which 6 trusses can unfold to a pillar-like structure shown at the zoomed plot in the top of Fig. \ref{fig:Hexagonal} (b), where for clarity only axial stresses are plotted. After that, the self-similar damage propagation process begins (CG). 
%For this reason, the simulation has not reached a point at which the crack propagation starts---a displacement of $\delta/h_{\mathrm{i}}=1.5$ has been applied at the DCB, which was still not enough to get the crack propagation to start. 
When the Kelvin interface undergoes the damage process, as in Fig. \ref{fig:Kelvin}, the first trusses to fail are not at the very front of the interface, but in the second column, where topology is different and the trusses are further constrained by their neighbours. 
%%%%%%%%%%%%%%%%%%%
%%%%%%%%%%%%%%%%%%%
Shortly after that, a large section of the interface fails at once which leads to snap-down in the force-displacement curve. This is followed by a damage propagation, which aligns with the fracture scaling law and it is similar to the one observed for\textit{ e.g.} the pillar interface.
Therefore, this leads to the consideration that the cells in their unloaded configuration have well defined finite in number symmetries, which are related to, or coincides with, the externally applied loading directions. However, due to fracture loading, these symmetries are broken, as the number of local degrees of freedom are different at the ``crack tip'' (or better described as ``damage front''), compared to other locations within the lattice~\cite{kane2014topological,zhang2018fracturing}. It is observed that certain topologies, those for which the arc length of the trusses between the joined surfaces exceeds the interface thickness, are likely and can be designed to unfold or rotate during damage propagation. This resembles the fracture phenomena in homogeneous thermoplastic materials where, within the damage zone, the macromolecular chains re-conform upon loading activating significant energy dissipation. 
%%%%%%%%%%%%%%%%%%
\subsection{Failure maps}
\begin{figure}[ht]
    \centering
    \includegraphics[width=8.5cm]{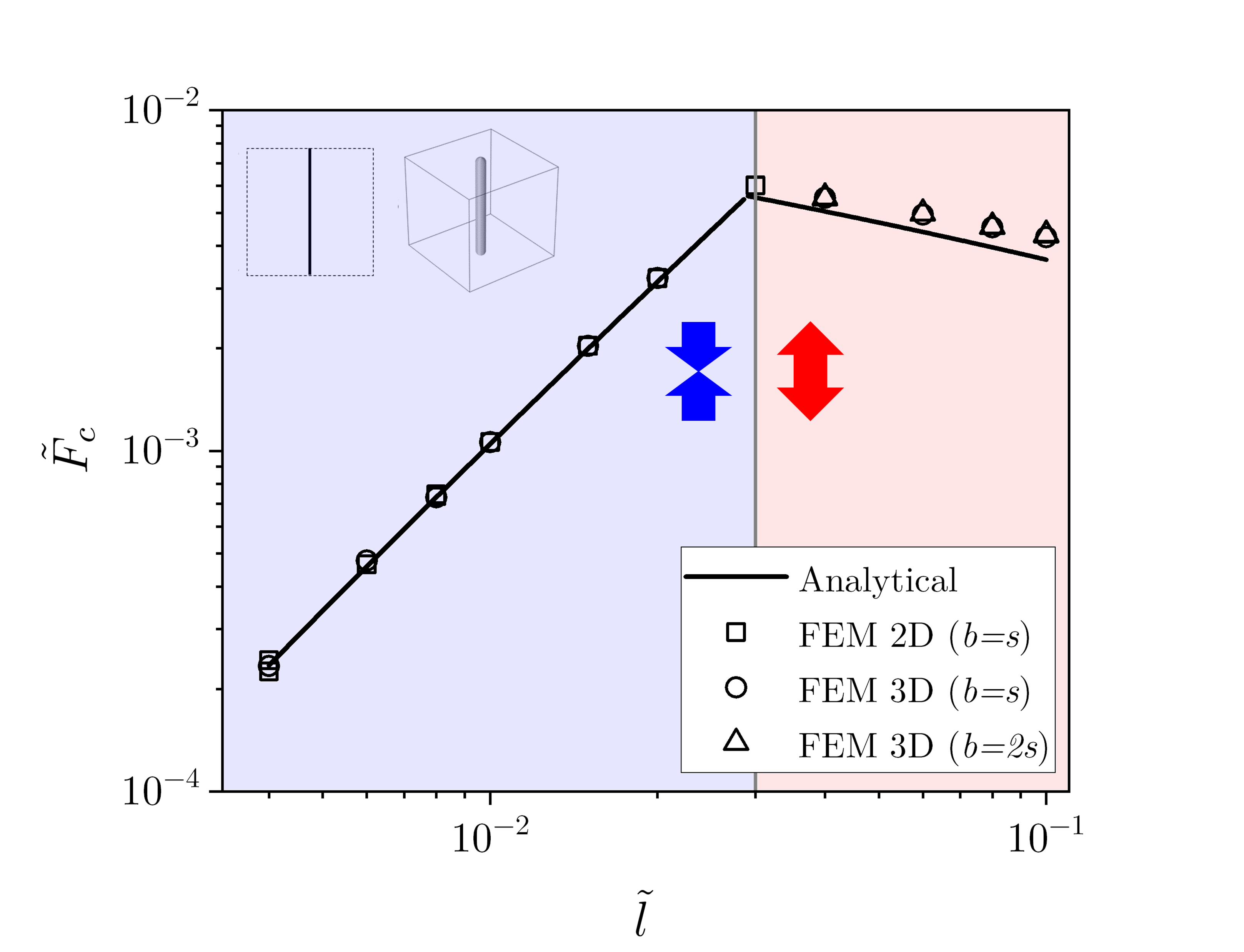}
    \caption{Failure map for the pillar interfaces. The region to the left (blue) shows compressive failure, whereas the region to the right (red) shows tensile failure.}
    \label{fig:Pillar_failure_map}
\end{figure}

%%%%%%%%%%%%%%%%%%
For each of the interface types considered here, we provide a failure map depicting the relationship between the normalised failure force $\tilde{F}_c$ and the aspect ratio $\tilde{l}$. These, which establish design spaces for ArchIs, enable further verification of results obtained through the analytical Eqs. (\ref{Eq:Fctilde_crit_t}) and (\ref{Eq:Fctilde_crit_c}). As previously, we focus on Case A BVP. For the pillar interface the definition of the critical force for the numerical studies is straightforward, as the force-displacement curve, Fig. \ref{fig:Pillar} is similar to what is known from a homogenised interface, with the difference of the stick-slip behaviour. The critical force is defined as the maximum force coinciding for pillar interfaces with the force under which failure of the first truss is recorded. The failure map for the pillar interface is presented in Fig. \ref{fig:Pillar_failure_map}, which shows a good agreement between the analytical and numerical results.
%%%%%%%%%%%%%%%%%%%
%%%%%%%%%%%%%%%%%%%

For the remaining interfaces, triangular, octet, hexagonal and Kelvin, the damage propagation is more complex, making it more ambiguous as for which force should be chosen as the critical force. Several different failure criteria have been considered, such as force corresponding to the failure of the first truss, the maximum force, and the force at which the damage propagation begins. All of these failure mechanisms are then collected for the failure maps for each failure mechanism.

\begin{figure}[ht]
    \centering
    \includegraphics[width=12.5cm]{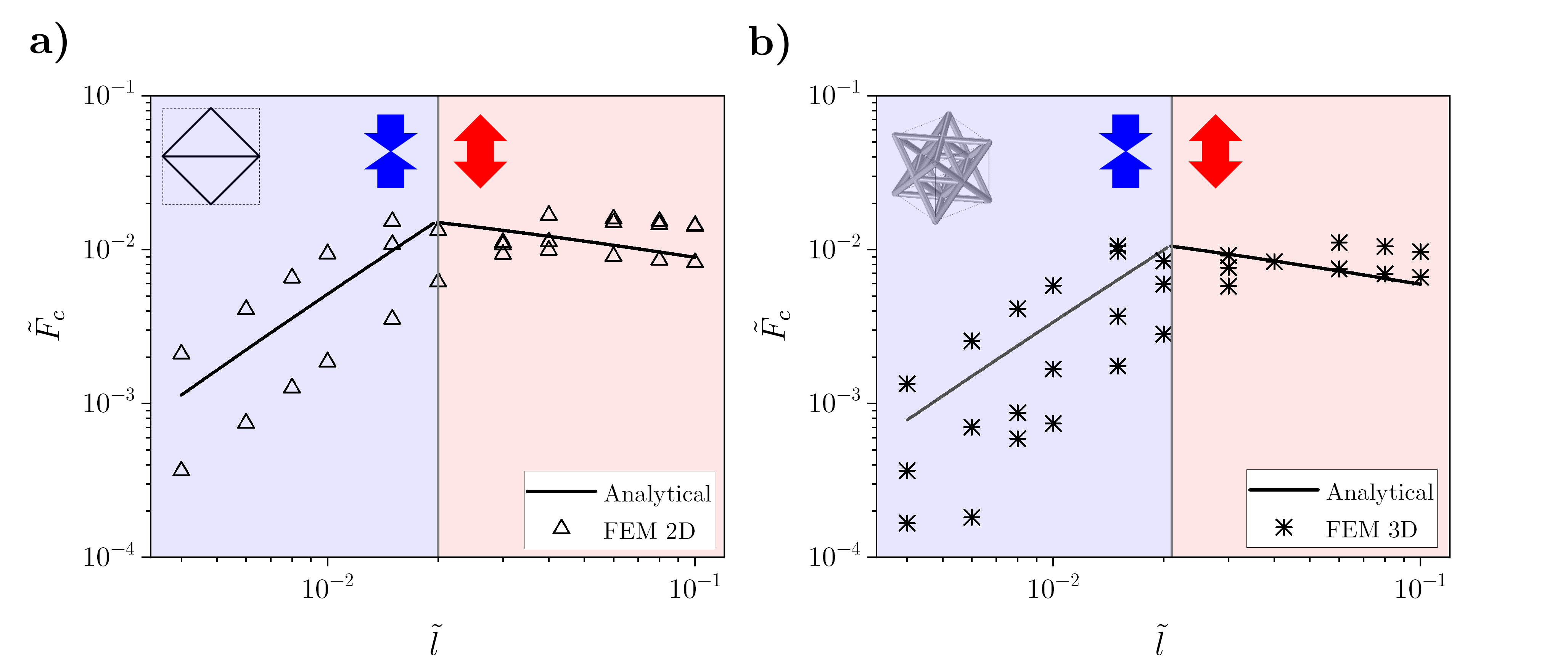}
    \caption{Failure maps for, (a) triangular, and, (b) octet cell interfaces. The region to the left (blue) shows compressive failure, whereas the region to the right (red) shows tensile failure.}
    \label{fig:FM_Triangular}
\end{figure}

Fig. \ref{fig:FM_Triangular} shows the failure maps for the two stretching dominated interfaces, triangular and octet. While the analytical and numerical results deviates quantitatively, it should the noted that the power low for the different failure mechanisms of the numerical results and the analytical formulation is in good agreement. The criterion chosen for the critical force obtained through FEM is not affecting the results qualitatively. 
%%%%%%%%%%%%%%%%%%%
%%%%%%%%%%%%%%%%%%%

\begin{figure}[h!]
    \centering
    \includegraphics[width=12.5cm]{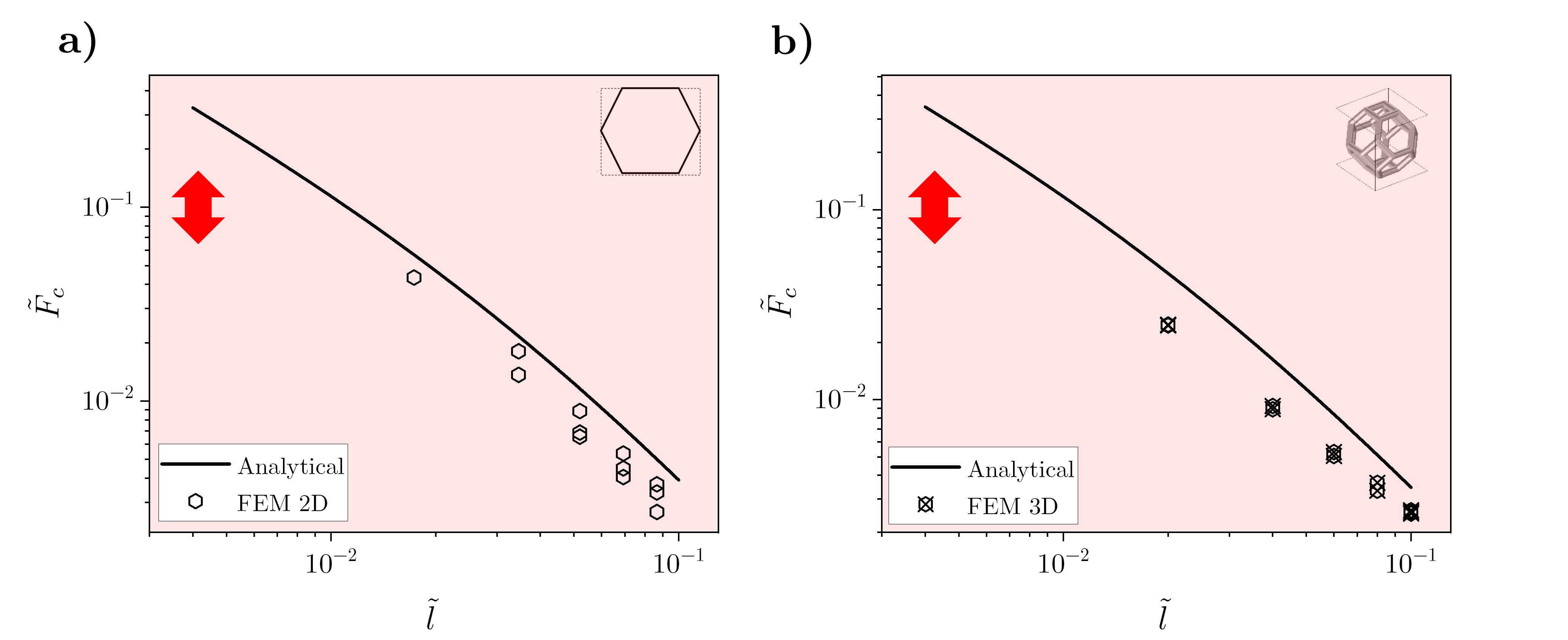}
    \caption{Failure maps for, (a) hexagonal, and, (b) Kelvin cell interfaces. The region to the left (blue) shows compressive failure, whereas the region to the right (red) shows tensile failure.}
    \label{fig:FM_Kelvin}
\end{figure}

The failure maps in Fig. \ref{fig:FM_Kelvin} correspond to the results for the bending-dominated interfaces---both hexagonal and Kelvin cells. As all the loads are leading to bending of individual trusses, these remain under tensile loading within the entire range of $\tilde{\lambda}$. For the bending-dominated cells, the critical strain-damage criterion is the same in both tension and compression. However, the displacement along the interface is much bigger within the tension zone than the compression leading to the trusses failing only due to tension. 
%%%%%%%%%%%%%%%%%%%
%%%%%%%%%%%%%%%%%%%

%%%%%%%%%%%%%%%%%%%%%%%%%%%%%%%%%%%%%%%%%%%%%
\section{Summary and Conclusions}
%%%%%%%%%%%%%%%%%%%%%%%%%%
%Behaviour of the ArchIs during fracture loading in two and three dimensions was investigated. Analytical and numerical models are presented to predict failure loads and analyse the results. Both models are found to be in a very good agreement for predicting initial material compliance and initial failure loads. Our main findings refer, however, to the post-failure-initiation stage. The different types of cell geometries, that are likely to fail under tensile loading in an unstable manner, lead to stable damage propagation under fracture loading. Such behaviour we attribute to the breakage of cell symmetry and new topological states near the damaged front. Characteristic of ArchIs (in contrast to architected materials) is the fact that both tensile and compressive stress zones are observed along the architected region, thus creating specific (related to the cell type) load profiles. The stick-slip damage propagation is then recorded and followed by step wise propagation of such specific patterns. We hope this work elucidates further the potential to exploit damage phenomena of architected materials to encode within the choice of MMs a characteristic toughening phenomena. We believe that there is still a need to further explore ArchIs in this context, as it seems an attractive solution for the every increasing applications of adhesive bonding within wind energy and marine industries.

\rv{The behaviour of ArchIs during fracture loading was investigated, yielding valuable insights into their ability to maintain structural integrity and enable stable damage propagation beyond the failure load. We employed tailored theoretical and numerical frameworks designed for assessing interfaces between two face materials, considering diverse microscopic geometries in both 2D (pillar, tetrahedron, hexagon) and 3D (pillar array, octet truss, Kelvin cell). Both frameworks showed excellent agreement in predicting structural compliance and critical failure loads. Figures \ref{fig:Pillar}--\ref{fig:Kelvin} showcase fracture performance in various architectural designs, emphasising crack propagation and elucidating toughening phenomena. However, our primary findings relate to the post-failure-initiation stage. Figures \ref{fig:Pillar_failure_map}--\ref{fig:FM_Kelvin} delve into damage initiation, highlighting the influence of topology, whether stretching or bending dominated. Despite our study's advancements, a notable limitation still exists in the absence of a robust fracture mechanics framework for accurately quantifying damage resistance parameters in architected materials, challenging traditional metrics like crack stress fields and the self-similarity of the fracture process zone.}

\rv{Different types of cell geometries, prone to unstable failure under tensile loading, lead to stable damage propagation during fracture loading, showcasing the potential for realising a fail-safe design. This behaviour is attributed to the breakage of cell symmetry and the emergence of new topological states near the damaged front. It was observed that crack propagation unfolds in distinct stages. Initially, under linear loading below the initial failure load, the material undergoes stress (EL). As the process advances, topologically preferable links emerge, bridging the face/beam materials and introducing substantial toughening effects (T). Subsequent to this, the failure of certain bridges occurs, giving rise to a quasi self-similar damage zone composed of both the remaining bridges and the leftover interface core. Finally, the crack undergoes growth, following the established pattern. This sequential progression sheds light on the dynamic evolution of material structure and the intricate mechanisms underlying crack propagation and toughening effects. Stick-slip damage propagation is recorded, followed by step-wise propagation of these specific patterns.}

\rv{The distinguishing feature of ArchIs lies in the concurrent presence of both tensile and compressive stress zones along the architected region, setting them apart from conventional architected materials. This simultaneous occurrence of tension and compression within the structure is a direct consequence of the unique design and arrangement of confinement. Unlike traditional architected materials where stress zones may be predominantly given by a $K$-field, ArchIs exhibit, in our case, a balanced distribution of tensile and compressive forces throughout their architected regions. This design characteristic leads to the development of specific load profiles closely tied to the cell type comprising the ArchIs. The interplay between tension and compression within the material not only influences its overall mechanical response but also contributes to the distinct behaviour observed during fracture loading, making ArchIs a compelling subject for understanding and harnessing the mechanical properties of MMs in various applications. Although this work does not exhaust all possible damage scenarios, as it studies only a few types of unit cells with perfect microstructures---an idealisation unlikely due to manufacturing flaws---we hope it further elucidates the potential to exploit the damage phenomena in architected materials and to incorporate characteristic toughening mechanisms within the selected MMs. Further exploration of ArchIs in this context is deemed necessary, offering an attractive solution for the ever-increasing applications of adhesive bonding to engineering solutions.}

\rv{\section*{CRediT authorship contribution statement}
M. L. S. Hedvard, M. A. Dias, and M. K. Budzik performed the research, analysed the methodologies and results, and wrote the paper. M. A. Dias and M. K. Budzik designed the research and acquired funds.}

% %%%%%%%%%%%%%%%%%%%%%%%%%%
\section*{Acknowledgements} 
M. K. Budzik would like to thank the Velux Foundations for support
under the Villum Experiment program (VIL50302). M. A. Dias would like to thank UKRI for support under the EPSRC Open Fellowship (EP/W019450/1).

%%%%%%%%%%%%%%%%%%%%%%%%%%
\appendix

%%%%%%%%%%%%%%%%%%%%%%%%%%%%%%%%%%%%%%%%%%%%%
%\section{Appendices}
%%%%%%%%%%%%%%%%%%%%%%%%%%

%%%%%%%%%%%%%%%%%%%%%
\section{Complementary calculations performed using Euler–Bernoulli Beam Theory}
\label{Appendix_A}
%%%%%%%%%%%%%%%%%%%%%
\subsection{Cases A and B}
For $x\leq a$ the Heaviside step function $H(x-a)=0$, thus eq. (\ref{Eq:EB_governing_eq}) can be reduced to: 
\begin{equation}
    \diff[4]{w}{x}=0.
    \label{A:Eq:EB_governing_x<a}
\end{equation}

By use of the BC from Table \ref{tab:BC_for_beams}, the shear force $Q_{xz}(x) = EIw'''(x)=F$ and moment $M(x) =w''(x) = Fx$ the solution to eq. (\ref{A:Eq:EB_governing_x<a}) is: 
\begin{equation}
    w_{x\leq a}(x)=\frac{F}{6EI}x^3+A_1 x + A_2
    \label{A:eq:sol_x<a_EB}
\end{equation}
where $A_1$ and $A_2$ are arbitrary constants. 

When $x>a$ the Heaviside step function $H(x-a)=1$, thereby the Winkler foundation is taken into account in eq. (\ref{Eq:EB_governing_eq}). Since the equation is homogeneous the complete solution is given by the complementary solution. The complimentary solution for an equation of this form is given by \citep{dillard2018review}:
\begin{equation}
    w_{x>a}(x) = A_3e^{- x/\lambda}\cos{ \frac{x}{\lambda}}+ A_4e^{- x/\lambda}\sin{ \frac{x}{\lambda}} + A_5e^{  x/\lambda}\cos{\frac{x}{\lambda}} + A_6e^{ x/\lambda}\sin{ \frac{x}{\lambda}}.
    \label{A:eq:sol_x>a_EB}
\end{equation}

The constants $A_2, A_3,...,A_6$ in eqs. (\ref{A:eq:sol_x<a_EB}) and (\ref{A:eq:sol_x>a_EB}) are determined using BC for case A and B from Table \ref{tab:BC_for_beams} and  continuity conditions where $w_{x\leq a}(a)=w_{x>a}(a)$,
$w'_{x\leq a}(a)=w'_{x>a}(a)$,  $w''_{x\leq a}(a)=w''_{x>a}(a)$, and
$w'''_{x\leq a}(a)=w'''_{x>a}(a)$. The two solutions for $w(x)$ is normalised by the relation $\tilde{w}\equiv w/h_{\mathrm{i}}$ and using the relations for the normalised force $\tilde{F}_{\mathrm{i}} \equiv F \lambda^3/(EI h_{\mathrm{i}})$, the normalised position $\tilde{x}\equiv x/\lambda$, the normalised crack length $\tilde{a}\equiv a/\lambda$, and the normalised length of the beam $\tilde{L}\equiv L/\lambda$.
The particular solution for case A and B are given in Table \ref{tab:wtilde_sol}.
%%%%%%%%%%%%%%%% Table 
\begin{table}[]
\caption{Solutions for the normalised beam deflections $\tilde{w}$ for the four different cases described in Table \ref{tab:BC_for_beams}.}
\centering
\begin{tabular}{c}
\hline\hline
\textbf{Case A} \\
\hline
$\begin{aligned}
 \tilde{w}(\tilde{x})=\frac{\tilde{F}_{\mathrm{i}}}{2} \times
    \begin{cases}
    & \text{for } \tilde{x} \leq \tilde{a}\\
    & \frac{1}{3} [1+\tilde{x}^3+2(1+\tilde{a})^3-3\tilde{x}(1+\tilde{a})^2] \\
    \quad\\
    & \text{for } \tilde{x} > \tilde{a}\\
    & e^{\tilde{a}-\tilde{x}} [(1+\tilde{a}) \cos(\tilde{a}-\tilde{x})+\tilde{a}\sin(\tilde{a}-\tilde{x})]\\
  \end{cases}
 \end{aligned}$ \\ \hline\hline
\textbf{Case B} \\
\hline
$
\begin{aligned}
\tilde{w}(\tilde{x}) = \frac{\Tilde{F}}{2}  \times
\begin{cases}
    \begin{aligned}
    & \text{for } \tilde{x} \leq \tilde{a}\\
    &\frac{1}{e^{4\Tilde{L}} + e^{4\Tilde{a}} + e^{2\Tilde{L}+2\Tilde{a}}[ 2\cos(2\Tilde{L}-2\Tilde{a})-4]}\\
    &\begin{aligned}
    \times \Big\{ &\frac{1}{3} e^{4\tilde{L}}[1+\tilde{x}^3+2(1+\tilde{a})^3-3\tilde{x}(1+a)^2]\\ &+e^{4\tilde{a}}[-1+\tilde{x}^3-2(1-\tilde{a})^3-3\tilde{x}(1-a)^2] \\
    &+2e^{2\tilde{L}+2\tilde{a}}[-2\tilde{x}^3+6\tilde{a}^2\tilde{x}-4\tilde{a}^3 + 12(\tilde{a}^2-\tilde{a}\tilde{x}-\frac{1}{2})\sin{(2\tilde{L}-\tilde{a})}\\
    & +(\tilde{x}^3+3\tilde{x}(1-\tilde{a}^2)-6\tilde{a}+2\tilde{a}^3) \cos{(2\tilde{L}-2\tilde{a})} ] \Big\}
    \end{aligned}
    \end{aligned}
    \\
    \quad\\
    \begin{aligned}
    & \text{for } \tilde{x} > \tilde{a}\\
    & \frac{B_1 e^{\tilde{x}}\sin{\tilde{x}} -B_2 e^{\tilde{x}}\cos{\tilde{x}} + B_3 e^{\tilde{x}}\cos{\tilde{x}} - B_4 e^{\tilde{x}}\sin{\tilde{x}}}{e^{4\Tilde{L}} + e^{4\Tilde{a}} + e^{2\Tilde{L}+2\Tilde{a}}[ 2\cos(2\Tilde{L}-2\Tilde{a})-4]}\\
    \end{aligned}
    \end{cases}
\end{aligned}
$
      \\
where   \\
$
\begin{aligned}
    B_1 = &e^{2\tilde{L}+\tilde{a}}[(\tilde{a}+1)\cos{(2\tilde{L}-\tilde{a})}-\tilde{a}\sin{(2\tilde{L}-\tilde{a})}-(2\tilde{a}+1)\cos{\tilde{a}}+\sin{\tilde{a}}]\\
    &+e^{3\tilde{a}} [(\tilde{a}-1)\sin{\tilde{a}}+\tilde{a}\cos{\tilde{a}}]
    \\
    B_2 = &e^{2\tilde{L}+\tilde{a}}
    [\tilde{a}\cos{(2\tilde{L}-\tilde{a})}+(\tilde{a}+1)\sin{(2\tilde{L}-\tilde{a})}-\cos{\tilde{a}}-(2\tilde{a}+1)\sin{\tilde{a}}]\\
    &+e^{3\tilde{a}} [(1-\tilde{a})\cos{\tilde{a}}+\tilde{a}\sin{\tilde{a}}]
    \\
    B_3 = & e^{2\tilde{L}+3\tilde{a}}
    [-\tilde{a}\cos{(2\tilde{L}-\tilde{a})}+(\tilde{a}-1)\sin{(2\tilde{L}-\tilde{a})}-\cos{\tilde{a}}+(1-2\tilde{a})\sin{\tilde{a}}]\\
    &+e^{4\tilde{L}+\tilde{a}} [(\tilde{a}+1)\cos{\tilde{a}}+\tilde{a}\sin{\tilde{a}}]
    \\
    B4 = & e^{2\tilde{L}+3\tilde{a}}
    [(\tilde{a}-1)\cos{(2\tilde{L}-\tilde{a})}+\tilde{a}\sin{(2\tilde{L}-\tilde{a})}+(1-2\tilde{a})\cos{\tilde{a}}+\sin{\tilde{a}}]\\
    &+e^{4\tilde{L}+\tilde{a}} [\tilde{a}\cos{\tilde{a}}-(\tilde{a}+1)\sin{\tilde{a}}]
\end{aligned}
$
        \\ \hline\hline
\textbf{Case C} \\
\hline
    $
    \begin{aligned}
    \tilde{w}(x) = \frac{\tilde{F}}{2}  \frac{[ -3(\tilde{L}+\tilde{a}) \tilde{x} + 2(\tilde{L}^2+\tilde{a}^2+\tilde{L}\tilde{a})]}{(\tilde{L}-\tilde{a})^3} 
     \end{aligned}$    \\ \hline \hline
\end{tabular}
\label{tab:wtilde_sol}
\end{table}
 %%%%%%%%%%%%%%%%%%%%%%%%%%%%%%%%%%%%%%%%%%%%%%%%%
\subsection{Rigid Beam - Case C}
When the interface section becomes sufficiently short, relative to the other dimensions and mechanical properties of the DCB, the beam will start to behave as a rigid beam. When this happens, the beam can be described with a linear equation:
\begin{equation}
w(x) = C_1 x + C_2    
\end{equation}
where $C_1$ and $C_2$ are constants that depends on the geometry and are here found using minimisation of the potential energy. 
The potential energy is given by: 
\begin{equation}
    \Pi = U-W 
\end{equation}
where $U$ is the strain energy and $W$ is the work done by the external loading. 

The work done by the external forces $F$ is given by:
\begin{equation}
    W = F w(x=0) = F C_2
\end{equation}
Since the beam is assumed to be rigid, the strain energy refers only to the energy stored in the interface. As the interface is assumed to be a linear elastic material, the strain energy can be expressed as:
\begin{equation}
    U = \frac{1}{2}\int \sigma_{ij} \epsilon_{ij} dV\\
\end{equation}
where $V$ refers to the volume of the interface. Within the Winkler foundation assumptions, the stress and strain component are found solely along the $z$-direction, whereas all the other components are zero. Thus, $\sigma_{ij} \epsilon_{ij}=\sigma_{zz} \epsilon_{zz}$. Using Hooke's law the strain energy can be found as:
%---
\begin{equation}
\begin{aligned}
U 
= \frac{1}{2}\int_a^L E_{\mathrm{i}} \epsilon_{zz}^2 dV
= \frac{1}{2}\int_a^L E_{\mathrm{i}} w(x)\frac{2}{h_{\mathrm{i}}}b dx
\end{aligned}
\end{equation}
%---
be aware that the $i$ in $E_{\mathrm{i}}$ here refers to the interface type. 

Minimisation of the potential energy is used to determine the constants $C_1$ and $C_2$, by setting:
\begin{equation}
    \frac{\partial\Pi}{\partial C_1}=\frac{\partial\Pi}{\partial C_2}=0.
\end{equation}
This gives: 
\begin{equation}
    \begin{aligned}
    C_1 = -\frac{3 F h_{\mathrm{i}} (L+a)}{E_{\mathrm{i}} b (L-a)^3}\\
    C_2 = \frac{2 F h_{\mathrm{i}} (L^2+a^2+La)}{E_{\mathrm{i}} b (L-a)^3}.
    \end{aligned}
\end{equation}
Then: 
\begin{equation}
    w(x) = \frac{F h_{\mathrm{i}}}{E_{\mathrm{i}} b (L-a)^3 } [ -3(L+a) x + 2(L^2+a^2+La)].
    \label{A:eq:w(x)_caseC}
\end{equation}

For comparison of Cases A, B and C, eq. (\ref{A:eq:w(x)_caseC}) is rewritten by using the normalised parameters found when solving cases A and B. The normalised solution for Case C is given in Table \ref{tab:wtilde_sol}.

%\bibliography{biball}

\end{document}